\documentclass[final,1p,times]{elsarticle}
\journal{Applied Mathematical Modelling}

\def\mysep{\sep}

\newtheorem{remark}{Remark}

\usepackage[colorlinks,bookmarksopen,bookmarksnumbered,citecolor=red,urlcolor=red,breaklinks=true]{hyperref}

\usepackage{showlabels}

\usepackage{amsmath}

\def\pp#1#2{\frac{\partial #1}{\partial #2}}

\def\bm#1{\boldsymbol{#1}}

\def\imath{\textrm{i}}

\def\diff{\mathrm{d}}

\usepackage{siunitx}
\NewDocumentCommand\numprint{m}{\num[round-mode = places]{#1}}
\NewDocumentCommand\nprounddigits{m}{\sisetup{round-precision = #1}}
\def\npproductsign#1{} 
\def\diff{\mathrm{d}}

\def\rot{\mathop{\mathrm{curl}}}

\def\reason#1{(\because~\text{#1})}

\def\vB{\bm{B}} 
\def\vx{\bm{x}}
\def\vs{\bm{s}}

\def\vV{\bm{V}}

\def\vP{\bm{P}}
\def\vd{\bm{d}}

\def\vA{\bm{A}} 

\def\Oee{O(\epsilon^2)}

\def\tilde#1{\widetilde{#1}}

\usepackage{pgfplots}
\usetikzlibrary{decorations.markings}

\def\supp{\mathop{\mathrm{supp}}} 

\def\Ncoil{N_{\rm coil}} 
\def\Npair{N_{\rm pair}} 
\def\Ncp{N} 
\def\ipair{\kappa}
\def\ip{\alpha_\ipair}
\def\jp{\beta_\ipair}

\usepackage[export]{adjustbox} 

\usepackage{ulem}
\usepackage{framed}

\usepackage{marginnote}

\def\mytitle{A shape optimisation of mutual inductances among coils}

\def\myabstract{%
  This paper introduces a shape optimisation framework for achieving desired mutual inductances (MIs) among coils in 3D space. Utilising a wire modelling approach, the coils are discretised using B-spline curves, with control points (CPs) serving as design variables. The key contribution is the derivation of the shape derivative of the objective function in terms of MIs, enabling the use of gradient-based quasi-Newton optimisation methods. A coil length constraint is also incorporated. The study demonstrates the effectiveness of the framework through numerical examples, validating the theoretical and numerical developments. This approach addresses the largely unexplored area of magnetostatic MI optimisation within the wire modelling framework, offering a computationally efficient alternative to finite element methods etc.
}%

\def\mykeywords{%
  Mutual inductance\mysep
  B-spline function\mysep
  Shape optimisation\mysep
  Shape derivative\mysep
  Biot--Savart law\mysep
  Nonlinear programming
}

\begin{document}

\begin{frontmatter}
  
  \title{\mytitle}
  
  \author[NU]{Toru Takahashi\corref{cor}}
  \ead{toru.takahashi@mae.nagoya-u.ac.jp}
  \author[NU]{Tatsuya Tokito}
  \author[NU]{Yi Cui}
  \author[NU]{Toshiro Matsumoto}
  \cortext[cor]{Corresponding author}
  \address[NU]{Department of Mechanical Systems Engineering, Nagoya University, Furo-cho, Chikusa-ku, Nagoya city, Aichi, 464-8603 Japan}
  
  \begin{abstract}
    \myabstract
  \end{abstract}
  
\end{frontmatter}

\section{Introduction}\label{s:intro}

Mutual inductance (MI) plays a pivotal role in a wide range of devices, including transformers, wireless power transfer systems, and near-field sensors. Enhancing the performance of these devices necessitates accurate analysis and optimisation of MIs among coils. Coils are typically modelled either as infinitely-thin wires carrying a prescribed current (scalar), or as finite-volume conductors with a current density (vector field) determined by Maxwell's equations. While the latter approach more accurately captures coil geometry, it entails significant computational complexity and cost due to the use of numerical methods like the finite element method (FEM).

In contrast, the simplified wire model offers a computationally efficient alternative. The Biot--Savart (BS) law, expressed as a contour integral, enables direct calculation of the magnetostatic field given the coil shape and current \cite{grover2004inductance}. This approach has been extensively used to investigate MIs in various coil configurations. For instance, Kim et al. \cite{chan2000simplified_neumanns_formula} computed the MI of spiral coils, Boissoles et al. derived the MI for bird-cage coils using integral forms and Gaussian quadrature, and Sonntag et al. \cite{sonntag2008implementation_of_the_neumann} examined a formula to calculate MI between planar printed circuit boards (PCBs). Akyel et al. \cite{akyel2009} studied the MI of circular and square coils based on Grover's formula \cite{grover2004inductance} and a numerical evaluation of the BS law. Joy et al. \cite{joy2014aacurate_computation_of_mutual} numerically evaluated the BS law to calculate the MI between two square coils in air when they are placed with various misalingnments; a similar investigation is found in the paper by Zhang et al.~\cite{zhang2018mutual_inductance_calculation}. Further research by Cheng et al. \cite{cheng2014a_new_analytical_calculation_of_the_mutual}, Paese et al. \cite{paese2015}, and Itoh et al. \cite{itoh2020empirical} explored MIs in various coil geometries, including spiral rectangular and toroidal-field coils.

Meanwhile, it should be noted that wire models are not always applicable, especially when dealing with skin effects and low-frequency time-varying electromagnetic fields. Numerical models (e.g., using finite elements with appropriate discretisations to manage the eddy current problem and the high field gradients near the conductor's surface \cite{seoane2019,miah2025}) are required in these instances to account for such effects.

Optimisation of MIs has been addressed using both bulk and wire modelling. Bulk modelling approaches, often employing commercial FEM solvers, have been applied in neuromodulation \cite{tian2022design_and_evaluation} and wireless power transfer \cite{yadav2025}. Wire modelling has seen more widespread use \cite{kiani2011,duan2012,krestovnikov2020,fadhel2023}, typically involving lumped-element models with electromagnetic and geometrical parameters as optimisation variables. Notably, these optimisation studies \cite{tian2022design_and_evaluation,yadav2025,kiani2011,duan2012,krestovnikov2020,fadhel2023} have primarily focused on frequency-domain dynamics.

However, magnetostatic MI optimisation within the wire modelling framework, the focus of this paper, remains largely unexplored. A related study by Hurwitz et al. \cite{hurwitz2025electromagneticcoiloptimizationreduced} utilised a Fourier expansion of coil geometry, treating the coefficients and current as design variables. These coefficients are analogous to the control points (CPs) used here. Nevertheless, their work did not optimise MIs directly.

This study builds upon the first author's recent work \cite{takahashi2024MRI} on gradient coil optimisation for MRI, which focused on generating linear magnetostatic fields. Here, the aim of this study is to optimise MIs to target values. To achieve this, coils are approximated with B-spline curves and their CPs are exploited as the design variables. Our primary contribution is the novel derivation of the shape derivative (or shape sensitivity) of MIs with respect to these B-spline control points. This critical development enables the use of powerful gradient-based quasi-Newton optimisation solvers to determine the CPs efficiently, providing a practical and computationally efficient alternative to more resource-intensive methods like finite element analysis for MI optimisation.

The paper is structured as follows: Section \ref{s:mutual} defines the optimisation problem and derives the shape derivative of the objective function. Section \ref{s:disc} details the discretisaion of coils and derivative with the B-spline functions. Section \ref{s:num} presents numerical examples to verify and validate the proposed optimisation method.
 
\section{Shape optimisation for mutual-inductances (MIs)}\label{s:mutual}

\subsection{Calculation of MIs}

\begin{figure}[hbt]
  \centering
  \includegraphics[width=.3\textwidth]{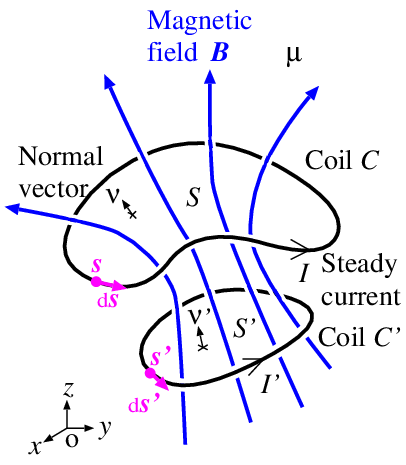}
  \caption{Schematic illustration of the magnetic flux $\Phi$ for a coil $C$ (receiver) due to the magnetostatic field $\vB$ induced by the steady current $I'$ running on another coil $C'$ (transmitter).}
  \label{fig:Phi}
\end{figure}

Let us consider two coils or closed curves, denoted by $C$ and $C'$, on which the steady currents $I$ and $I'$ are running, respectively, in the free space with the magnetic permeability $\mu$ (Figure \ref{fig:Phi}). Here, $C$ and $C'$ correspond to the receiver and transmitter, respectively, in terms of wireless power transmitting. Then, the magnetostatic field $\vB$ at a point $\vx$ due to $C'$ is represented by the BS law, i.e.
\begin{align}
  \vB(\vx) = \frac{\mu I'}{4\pi}\int_{C'}\frac{\diff\vs'\times(\vx-\vs')}{|\vx-\vs'|^3}.
  \label{eq:B}
\end{align}
Here, SI units are employed throughout this paper. The corresponding vector potential $\vA$ is given by
\begin{align}
  \vA(\vx) = \frac{\mu I'}{4\pi}\int_{C'}\frac{\diff\vs'}{|\vx-\vs'|},
  \label{eq:A}
\end{align}
which satisfies $\vB = \rot\vA$. Subsequently, let us denote any open (virtual) surface, whose boundary agrees with $C$, by $S$, where $\partial S\equiv C$ holds. Then, the magnetic flux $\Phi$ through $S$ is defined and calculated as follows:
\begin{align}
  \Phi
  :=\int_S\vB\cdot\bm{\nu}\ \diff S
  =\int_S\rot\vA\cdot\bm{\nu}\ \diff S
  =\int_C\vA\cdot\diff\vs
  =\frac{\mu I'}{4\pi}\int_C\int_{C'}\frac{\diff\vs\cdot\diff\vs'}{|\vs-\vs'|},
  \label{eq:Phi}
\end{align}
where $\bm{\nu}$ denotes the unit normal vector to $S$ and Stokes's theorem was used to convert the surface integral (over $S$) to the contour integral. It is known that the value of $\Phi$ does not depend on the choice of $S$. Then, the MI, denoted by $M$, between $C$ and $C'$ is defined by dividing $\Phi$ by $I'$ as follows:
\begin{align}
  M := \frac{\Phi}{I'} = \frac{\mu}{4\pi}\int_C\int_{C'}\frac{\diff\vs\cdot\diff\vs'}{|\vs-\vs'|},
  \label{eq:M}
\end{align}
which is well known as Neumann formula. It should be noted that the value of $M$ is maintained even if the roles of the receiver and transmitter is swapped.

\begin{remark} 

The BS law in (\ref{eq:B}) provides a valid approximation for transient cases within low-frequency regimes, i.e. quasi-static approximation, where the displacement current is negligible \cite{jackson1999classical}. Consequently, the SO framework presented herein can be extended to address such transient problems, although this paper focuses exclusively on the magnetostatic case. However, it is important to note that the assumption of treating a conducting wire as a (volume-less) curve, inherent in this framework, cannot account for physical phenomena such as the skin effect (as mentioned in the introduction). This limitation implies that the model may not accurately represent scenarios where such volumetric effects are significant.

\end{remark}

\subsection{Shape optimisation problem}\label{eq:mutual_problem}

Let us find shapes of the coils $C$ and $C'$ that can minimise the following objective function $J$ in terms of the MI $M$ in (\ref{eq:M}):
\begin{align}
  J:=\frac{1}{2}(M-\overline{M})^2,
  \label{eq:J}
\end{align}
where $\overline{M}$ denotes the target MI given \textit{a priori}. In addition, geometric constraints are considered to refrain the coils from becoming too large or small. The details of such constraints are mentioned in Section~\ref{s:constraints} after establishing how the shapes of coils are theoretically and numerically treated in this study.

\begin{figure}[hbt]
  \centering
  \includegraphics[width=.4\textwidth]{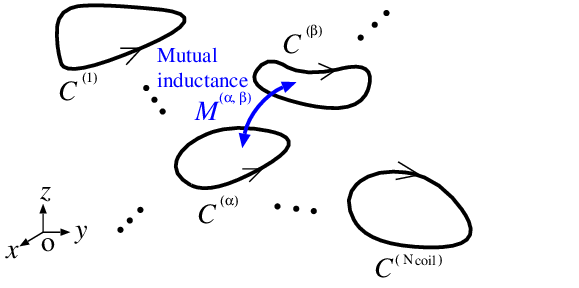}
  \caption{Generalised configuration of the shape optimisation for $\Ncoil$ coils.}
  \label{fig:multi}
\end{figure}

\begin{remark}{Generalisation to arbitrary number of coils.}\label{remark:gen}
  As illustrated in Figure~\ref{fig:multi}, let us consider $\Ncoil$ coils $C^{(1)}$, $\ldots$, $C^{(\Ncoil)}$. In what follows, the symbol with the superscript `$(\alpha)$' is associated with the $\alpha$th coil, i.e. $C^{(\alpha)}$. Then, let us find shapes of the coils that can maximise the following generalised objective function, which is also denoted by $J$, in terms of the MIs of $\Npair$ pairs of coils:
  \begin{align}
    J:=\frac{1}{2}\sum_{\ipair=1}^{\Npair} \left(M^{(\ip,\jp)}-\overline{M}^{(\ip,\jp)}\right)^2,
    \label{eq:Jgen}
  \end{align}
where $\ip\in[1,\Ncoil]$ and $\jp\in[1,\Ncoil]$ represent the indices of coils for the $\ipair$th pair, denoted by $(\ip,\jp)$, where $\ip\ne\jp$ is supposed. Also, $M^{(\ip,\jp)}$ ($\equiv M^{(\jp,\ip)}$) denotes the MI between $C^{(\ip)}$ and $C^{(\jp)}$ and can be calculated according to (\ref{eq:M}). The choice of $\Npair$ pairs is arbitrary and is determined by the specific optimisation problem being considered.
\end{remark}

\subsection{Shape derivative of $M$}

The theoretical highlight of this study is to derive the shape derivative of $J$ in (\ref{eq:J}). The derivation can be carried out by calculating the shape derivative of $M$ in (\ref{eq:M}) and then utilising the differential formula for composite function.

First, following the previous study \cite{takahashi2024MRI}, the shape derivative of $M$ in (\ref{eq:M}), denoted by $D$, is derived. To this end, suppose that any points $\vs$ and $\vs'$ on $C$ and $C'$ are perturbed to $\tilde{\vs}$ and $\tilde{\vs}'$ by infinitesimal perturbations (translations) $\epsilon\vV(\vs)$ and $\epsilon\vV'(\vs')$, respectively, meaning
\begin{align}
  \tilde{\vs}=\vs+\epsilon\vV(\vs),\quad
  \tilde{\vs}'=\vs'+\epsilon\vV'(\vs'),
  \label{eq:tilde s}
\end{align}
where the vectors $\vV(\vs)$ and $\vV'(\vs')$ are predefined at any points $\vs$ and $\vs'$ on $C$ and $C'$, respectively, and $\epsilon$ is an infinitesimally small positive number. Accordingly, the MI $M$ in (\ref{eq:M}) is perturbed to $\tilde{M}$ as follows:
\begin{align*}
  \tilde{M}
  &= \frac{\mu}{4\pi}\int_{\tilde{C}}\int_{\tilde{C}'}\frac{\diff\tilde{\vs}\cdot\diff\tilde{\vs}'}{|\tilde{\vs}-\tilde{\vs}'|} \nonumber\\
  &= \frac{\mu}{4\pi}\int_C\int_{C'}\frac{(\diff\vs+\epsilon\diff\vV)\cdot(\diff\vs'+\epsilon\diff\vV')}{|\vs+\epsilon\vV-\vs'-\epsilon\vV'|} \qquad\reason{Eq.~(\ref{eq:tilde s})} \nonumber\\
  &= \frac{\mu}{4\pi}\int_C\int_{C'}\frac{\diff\vs\cdot\diff\vs'+\epsilon(\diff\vs\cdot\diff\vV'+\diff\vV\cdot\diff\vs')+\epsilon^2\diff\vV\cdot\diff\vV'}{|\vs-\vs'+\epsilon(\vV-\vV')|}.
\end{align*}
Moreover, the inverse of the denominator in the above integrand can be expanded as
\begin{align*}
  \frac{1}{|\vs-\vs'+\epsilon(\vV-\vV')|}
  &= \frac{1}{\sqrt{|\vs-\vs'|^2+2\epsilon(\vs-\vs')\cdot(\vV-\vV')+\epsilon^2|\vV-\vV'|^2}}\nonumber\\
  &= \frac{1}{|\vs-\vs'|}\frac{1}{\sqrt{1+2\epsilon\frac{(\vs-\vs')\cdot(\vV-\vV')}{|\vs-\vs'|^2}+\epsilon^2\frac{|\vV-\vV'|^2}{|\vs-\vs'|^2}}}\nonumber\\
  &= \frac{1}{|\vs-\vs'|}\left(1-\epsilon\frac{(\vs-\vs')\cdot(\vV-\vV')}{|\vs-\vs'|^2}+\Oee\right)\quad(\epsilon\to 0).
\end{align*}
For brevity, the condition ``$\epsilon\to 0$'' under which the big-O notation $\Oee$ holds will be omitted hereafter.
Hence, one gets
\begin{align}
  \tilde{M}
  =& \frac{\mu}{4\pi}\int_C\int_{C'} \frac{1}{|\vs-\vs'|}\left(1-\epsilon\frac{(\vs-\vs')\cdot(\vV-\vV')}{|\vs-\vs'|^2}+\Oee\right)\left(\diff\vs\cdot\diff\vs'+\epsilon(\diff\vs\cdot\diff\vV'+\diff\vV\cdot\diff\vs')+\Oee\right) \nonumber\\
  =& \frac{\mu}{4\pi}\int_C\int_{C'} \frac{\diff\vs\cdot\diff\vs'}{|\vs-\vs'|}+\epsilon\frac{\mu}{4\pi}\int_C\int_{C'}\frac{1}{|\vs-\vs'|}\left[-\frac{(\vs-\vs')\cdot(\vV-\vV')}{|\vs-\vs'|^2}\diff\vs\cdot\diff\vs'+\diff\vs\cdot\diff\vV'+\diff\vV\cdot\diff\vs'\right]+\Oee \nonumber\\
  =& M+\epsilon\frac{\mu}{4\pi}\int_C\int_{C'}\left[\frac{\diff\vs\cdot\diff\vV'+\diff\vV\cdot\diff\vs'}{|\vs-\vs'|}-\frac{(\vs-\vs')\cdot(\vV-\vV')}{|\vs-\vs'|^3}\diff\vs\cdot\diff\vs'\right]+\Oee.
  \label{eq:tilde M}
\end{align}
From this, the shape derivative $D$ of $M$ is defined and calculated as follows:
\begin{align}
  D
  :=\lim_{\epsilon\to 0}\frac{\tilde{M}-M}{\epsilon}
  =\frac{\mu}{4\pi}\int_C\int_{C'}\left[\frac{\diff\vs\cdot\diff\vV'+\diff\vV\cdot\diff\vs'}{|\vs-\vs'|}-\frac{(\vs-\vs')\cdot(\vV-\vV')}{|\vs-\vs'|^3}\diff\vs\cdot\diff\vs'\right].
  \label{eq:D}
\end{align}

\subsection{Shape derivative of $J$}

Successively, the shape derivative of $J$ in (\ref{eq:J}) can be calculated by considering its perturbation. Specifically, the perturbed objective function, denoted by $\tilde{J}$, is expanded as follows:
\begin{align}
  \tilde{J}
  :=\frac{1}{2}(\tilde{M}-\overline{M})^2
  =\frac{1}{2}(\tilde{M}-M+M-\overline{M})^2
  =\frac{1}{2}(\tilde{M}-M)^2+(M-\overline{M})(\tilde{M}-M)+J.
  \label{eq:tilde J}
\end{align}
Therefore, the shape derivative of $J$, denoted by $G$, is calculated as 
\begin{align}
  G
  :=\lim_{\epsilon\to 0}\frac{\tilde{J}-J}{\epsilon}
  =\lim_{\epsilon\to 0}\left[ \frac{1}{2}\left(\frac{\tilde{M}-M}{\epsilon}\right)^2\epsilon +(M-\overline{M})\frac{\tilde{M}-M}{\epsilon} \right]
  =(M-\overline{M})D, 
  \label{eq:G}
\end{align}
where $D$ in (\ref{eq:D}) was used. This can be interpreted as the differentiation of the composite function $J=J(M)$ with respect to $M$.

\section{Discretisation with the B-spline functions}\label{s:disc}

According to the previous study~\cite{takahashi2024MRI}, the coil curves are approximated with the B-spline curves. As a result, the present SO problem is reduced to a discrete form where nonlinear-programming (NLP) solvers can be applied.

\subsection{Closed B-spline curves}

A closed curve $C$ is represented with the B-spline functions as follows:
\begin{align}
  \vs(t)=\sum_{m=0}^{N-1} R_m^p(t)\vP_m,
  \label{eq:s with R}
\end{align}
where the vectors $\vP_0$, $\ldots$, $\vP_{N-1}$ denote the control points (CPs) given in accordance with a desired shape of $C$ (before its SO). Also, the functions $R_0^p$, $\ldots$, $R_{N-1}^p$ denote the periodic B-spline functions of degree $p$ ($\ge 1$) \footnote{The superscript $p$ in $R_m^p$ does not denote a power. No power of B-spline functions is used in this paper.} and associated with non-overlapping $N+1$ knots, i.e. $\{t_0,t_1,\ldots,t_N\}$ where $0=t_0<t_1<\cdots<t_N=1$ holds. Correspondingly to the knots, the $k$th interval $I_k$ is defined by $[t_k,t_{k+1}]$, where $k=0,\ldots,N-1$. Then, the periodic B-spline function $R^p_m$ (where $m=0,\ldots,N-1$) is defined as follows:
\begin{align*}
  R^p_m(t):=&\begin{cases}
    N^p_0\left(\frac{t-t_m}{\Delta}\right) & \text{if $t\in I_{m\% N}$}\\
    N^p_1\left(\frac{t-t_m}{\Delta}\right) & \text{if $t\in I_{(m+1)\% N}$}\\
    \quad\vdots & \quad\vdots\\
    N^p_p\left(\frac{t-t_m}{\Delta}\right) & \text{if $t\in I_{(m+p)\% N}$}\\
    0 & \text{otherwise}
  \end{cases}
  \quad\text{for $t\in[t_0,t_N]$},
\end{align*}
where $N^p_k$ is the associated $p$th-order polynomial. Also, the symbol \% denotes the modulo operator: $n \% N$ denotes the remainder of $n$ divided by $N$. As a result, regarding the last $p$ functions, i.e. $R^p_{N-p}$, $\ldots$, $R^p_{N-1}$, their supports are split to two groups of intervals. Such splitting is necessary to satisfy the partition of unity (POU), i.e. $\sum_{m=0}^{N-1}R_m^p(t)=1$ for any $t\in[t_0,t_N]$. Specifically, the support of $R^p_m(t)$, denoted by $\supp(R^p_m)$, is consisting of $p+1$ intervals and represented as follows:
\begin{align}
  \supp(R^p_m)
  &= \begin{cases}
    \cup_{k=m}^{m+p}I_k & \text{for $m=0,\ldots,N-p-1$}\\
    \left(\cup_{k=0}^{m-N+p} I_k \right) \cup \left( \cup_{k=m}^{N-1}I_k \right) & \text{for $m=N-p,\ldots,N-1$}
  \end{cases}\nonumber\\
  &= \cup_{k=m}^{(m+p)\%N} I_k \quad \text{for $m=0,\ldots,N-1$}.
  \label{eq:supp}
\end{align}
This means that $R^p_{N-p}$, $\ldots$, $R^p_{N-1}$ has split-intervals. So, it should be noted that, if the starting index $n$ is \textit{larger} than the ending index $(n+p)\%N$, the union $\cup_{k=m}^{(m+p)\%N}$ in the most RHS of (\ref{eq:supp}) should be interpreted as two unions $\cup_{k=0}^{(m+p)\%N}$ and $\cup_{k=m}^{N-1}$.

Similarly to $C$, the other coil $C'$ is presented with a closed B-spline curve of degree $p'$ associated with $N'$ CPs.

\begin{figure}[hbt]
  \centering
  \includegraphics[width=.8\textwidth]{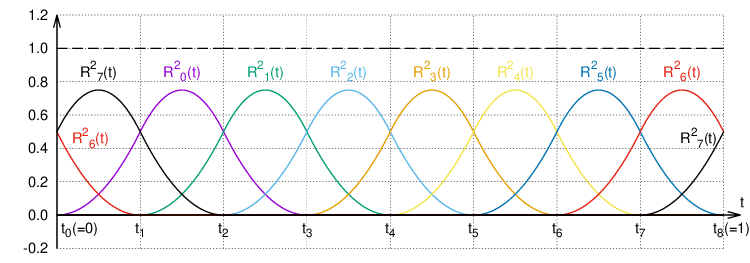}
  \caption{Example of the periodic B-spline functions of degree 2 when using $N=8$, viz. $R_0^2$, $\ldots$, $R_7^2$, and the uniform knots such as $t_m=m/N$. The dashed line is the summation of all the functions and one, meaning the partition of unity is satisfied.}
  \label{fig:bspline-p2N8}
\end{figure}

Figure~\ref{fig:bspline-p2N8} displays an example of the the periodic B-spline functions of degree $2$ ($=p$) with using $N=8$ and the uniform knots, i.e. 
\begin{align*}
  t_m:=m\Delta\quad\text{for $m=0,\ldots,N$},
\end{align*}
where $\Delta:=\frac{1}{N}$ denotes the knot span. Also, the corresponding polynomial functions $N^2_k$ (where $k=0$, $1$, and $2$) are given as
\begin{align*}
  N^2_0(x):=\frac{x^2}{2},\qquad
  N^2_1(x):=\frac{-2x^2+6x-3}{2},\qquad
  N^2_2(x):=\frac{(x-3)^2}{2}.
\end{align*}
In the figure, it is observed that the last two functions $R^2_6$ and $R^2_7$ are split in two portions. Regarding $R^2_7$, a union $\cup_{k=7}^1$ in the support is actually interpreted as $\cup_{k=0}^1$ and $\cup_{k=7}^{N-1}$. In addition, the summation of the all the functions (drawn with the dashed line) is one for any $t$ in $[t_0,t_N]$, which means that the POU is satisfied. 

\subsection{Discretisation of $M$, $\tilde{M}$, and $\tilde{J}$}

First, the discretisation of $M$ in (\ref{eq:M}) with the closed B-spline curve is described. The substitution of $\vs$ in (\ref{eq:s with R}) and the corresponding representation for $\vs'$, which is based on $N'$ CPs of $C'$, into $M$ yields
\begin{align}
  M
  &= \frac{\mu}{4\pi}\int_C\int_{C'}\frac{\sum_{a=0}^{N-1}\frac{\diff R_a^p(t)}{\diff t}\diff t\vP_a\cdot\sum_{b=0}^{N'-1}\frac{\diff R_b^{p'}(t')}{\diff t'}\diff t'\vP'_b}{|\vs(t)-\vs'(t')|} \nonumber\\
  &= \frac{\mu}{4\pi}\sum_{a=0}^{N-1}\sum_{b=0}^{N'-1}\int_C\int_{C'}\frac{\dot{R}^p_a(t)\dot{R}^{p'}_b(t')}{|\vs(t)-\vs'(t')|}\diff t\diff t' \vP_a\cdot\vP'_b && \reason{Exchange of summations and integrals}\nonumber\\
  &= \sum_{a=0}^{N-1}\sum_{b=0}^{N'-1} \underbrace{\frac{\mu}{4\pi} \sum_{k=a}^{(a+p)\%N}\sum_{l=b}^{(b+p)\%N'}\int_{I_k}\dot{R}^p_a(t) \left(\int_{I_l}\frac{\dot{R}^{p'}_b(t')}{|\vs(t)-\vs'(t')|}\diff t' \right) \diff t}_{=:m_{a,b}} \vP_a\cdot\vP'_b && \reason{$\int_C\diff t=\sum_{k=a}^{(a+p)\%N}\int_{I_k}\diff t$ holds from Eq.~(\ref{eq:supp})} \nonumber\\
  &= \sum_{a=0}^{N-1}\sum_{b=0}^{N'-1} m_{a,b} \vP_a\cdot\vP'_b,
  \label{eq:M with R}
\end{align}
where the dot $\dot{(\ )}$ denotes the differentiation with respect to the curve parameter $t$ or $t'$. It should be noted that the points $\bm{s}(t)$ and $\bm{s}'(t')$ in the above and following expressions can be computed from the B-spline approximation in (\ref{eq:s with R}).

Next, let us discretise the perturbation $\tilde{M}$ in (\ref{eq:tilde M}). Similarly to (\ref{eq:s with R}), the perturbed point $\tilde{\vs}=\tilde{\vs}(t)$ on $\tilde{C}$ is generated by the perturbed CPs, i.e. $\tilde{\vP}_0$, $\ldots$, $\tilde{\vP}_{N-1}$, as follows:
\begin{align}
  \tilde{\vs}(t)=\sum_{m=0}^{N-1} R_m^p(t)\tilde{\vP}_m.
  \label{eq:tilde s with R}
\end{align}
From (\ref{eq:s with R}) and (\ref{eq:tilde s with R}), one has
\begin{align}
  \tilde{\vs}(t)-\vs(t)
  =\sum_{m=0}^{N-1} R_m^p(t)(\tilde{\vP}_m-\vP_m)
  =\sum_{m=0}^{N-1} R_m^p(t)\delta\vP_m,
\end{align}
where $\delta\vP_m:=\tilde{\vP}_m-\vP_m$ denotes the perturbation of $\vP_m$. This is compared with (\ref{eq:tilde s}) to obtain
\begin{align}
  \epsilon\vV(\vs)=\sum_{m=0}^{N-1} R_m^p(t)\delta{\vP}_m.
  \label{eq:eV}
\end{align}
The similar relationship can be established for $C'$, i.e.
\begin{align}
  \epsilon\vV'(\vs')=\sum_{n=0}^{N'-1} R_n^{p'}(t')\delta{\vP}'_n.
  \label{eq:eV'}
\end{align}
These relationships enable to rewrite $\tilde{M}$ as follows:
\begin{align}
  \tilde{M}
  &= M + \epsilon D + \Oee \qquad\reason{Eqs.~(\ref{eq:tilde M}) and (\ref{eq:D})} \nonumber\\
  &= M + \frac{\mu}{4\pi}\int_C\int_{C'}\left[\frac{\diff\vs\cdot\diff(\epsilon\vV')+\diff(\epsilon\vV)\cdot\diff\vs'}{|\vs-\vs'|}-\frac{(\vs-\vs')\cdot(\epsilon\vV-\epsilon\vV')}{|\vs-\vs'|^3}\diff\vs\cdot\diff\vs'\right]+\Oee \nonumber\\
  &= M + \frac{\mu}{4\pi}\int_C\int_{C'}\Biggl[\sum_{n=0}^{N'-1}\frac{\dot{R}^{p'}_n(t')}{|\vs-\vs'|}\dot{\vs}\cdot\delta\vP'_n+\sum_{m=0}^{N-1}\frac{\dot{R}^{p}_m(t)}{|\vs-\vs'|}\dot{\vs}'\cdot\delta\vP_m\nonumber\\
    &-\sum_{m=0}^{N-1} \frac{R^{p}_m(t)\dot{\vs}\cdot\dot{\vs}'}{|\vs-\vs'|^3}(\vs-\vs')\cdot\delta\vP_m
    +\sum_{n=0}^{N'-1} \frac{R^{p'}_n(t')\dot{\vs}\cdot\dot{\vs}'}{|\vs-\vs'|^3}(\vs-\vs')\cdot\delta\vP'_n\Biggr]\diff t\diff t' +\Oee \nonumber\\
  &\qquad\qquad\qquad\qquad\qquad\qquad\qquad\qquad\qquad\qquad\reason{Eqs.~(\ref{eq:eV}) and (\ref{eq:eV'})} \nonumber\\
  &= M
  + \sum_{m=0}^{N-1}\underbrace{\frac{\mu}{4\pi}\int_C\int_{C'}\Biggl[\frac{\dot{R}^{p}_m(t)}{|\vs-\vs'|}\dot{\vs}'
    - \frac{R^{p}_m(t)\dot{\vs}\cdot\dot{\vs}'}{|\vs-\vs'|^3}(\vs-\vs')\Biggr]\diff t\diff t'}_{=:\vd_m}\cdot\delta\vP_m +\Oee \nonumber\\
  &+ \sum_{n=0}^{N'-1}\underbrace{\frac{\mu}{4\pi}\int_C\int_{C'}\Biggl[\frac{\dot{R}^{p'}_n(t')}{|\vs-\vs'|}\dot{\vs}
      + \frac{R^{p'}_n(t')\dot{\vs}\cdot\dot{\vs}'}{|\vs-\vs'|^3}(\vs-\vs')\Biggr]\diff t\diff t'}_{=:\vd'_n}\cdot\delta\vP'_n +\Oee \nonumber\\
  &= M + \sum_{m=0}^{N-1} \vd_m \cdot\delta\vP_m + \sum_{n=0}^{N'-1}\vd'_n\cdot\delta\vP'_n +\Oee,
  \label{eq:tilde M with R}
\end{align}
where the vectors $\vd_m$ and $\vd'_n$ represent the shape sensitivities of $M$ with respect to the $m$th CP $\vP_m$ of $C$ and the $n$th CP $\vP'_n$ of $C'$, respectively. Moreover, since the support of $R^p_m$ is $\cup_{k=m}^{(m+p)\% N} I_k$, the integral over $C$ in $\vd_m$ can be rewritten with the integrals over $p+1$ intervals. This is true for the integral over $C'$ in $\vd'_n$. As a result, the sensitivities can be represented as follows:
\begin{subequations}
  \begin{align}
    \vd_m
    &=\frac{\mu}{4\pi}\sum_{k=m}^{(m+p)\% N}\int_{I_k}\int_{C'}\Biggl[ \frac{\dot{R}^{p}_m(t)}{|\vs(t)-\vs'(t')|}\dot{\vs}'(t') - \frac{R^{p}_m(t)\dot{\vs}(t)\cdot\dot{\vs}'(t')}{|\vs(t)-\vs'(t')|^3}(\vs(t)-\vs'(t')) \Biggr]\diff t\diff t', \label{eq:vd1}\\
    \vd'_n
    &=\frac{\mu}{4\pi}\sum_{l=n}^{(n+p')\% N'}\int_{I_l}\int_C \Biggl[\frac{\dot{R}^{p'}_n(t')}{|\vs(t)-\vs'(t')|}\dot{\vs}(t) + \frac{R^{p'}_n(t')\dot{\vs}(t)\cdot\dot{\vs}'(t')}{|\vs(t)-\vs'(t')|^3}(\vs(t)-\vs'(t'))\Biggr]\diff t\diff t'\nonumber\\
    &=\frac{\mu}{4\pi}\sum_{l=n}^{(n+p')\% N'}\int_{I_l}\int_C \Biggl[\frac{\dot{R}^{p'}_n(t')}{|\vs'(t')-\vs(t)|}\dot{\vs}(t) - \frac{R^{p'}_n(t')\dot{\vs}'(t')\cdot\dot{\vs}(t)}{|\vs'(t')-\vs(t)|^3}(\vs'(t')-\vs(t))\Biggr]\diff t'\diff t.
    \label{eq:vd2}
  \end{align}%
  \label{eq:vd}%
\end{subequations}%

Last, the perturbation $\tilde{J}$ in (\ref{eq:tilde J}) is considered. Since
\begin{align*}
  \tilde{J}
  = J+\epsilon G + \Oee
  = J+(M-\overline{M})(\epsilon D) + \Oee
\end{align*}
holds from (\ref{eq:tilde J}) and (\ref{eq:G}), $\tilde{J}$  can be expressed as
\begin{align}
  \tilde{J}
  &= J + \left( \sum_{a=0}^{N-1}\sum_{b=0}^{N'-1} m_{a,b} \vP_a\cdot\vP'_b - \overline{M} \right) \left( \sum_{m=0}^{N-1} \vd_m \cdot\delta\vP_m + \sum_{n=0}^{N'-1}\vd'_n\cdot\delta\vP'_n \right) + \Oee,
  \label{eq:tilde J with R}
\end{align}
where the discretised $M$ in (\ref{eq:M with R}) and $\epsilon D$ in (\ref{eq:tilde M with R}) were used.

\begin{remark}{Generalisation of $\tilde{J}$.}
  The expression $\tilde{J}$ in (\ref{eq:tilde J with R}) can be enhanced to the generalised objective function $J$ mentioned in Remark~\ref{remark:gen} as
  \begin{align}
    \tilde{J}
    &= J + \sum_{\ipair=1}^{\Npair} \left( \sum_{a=0}^{\Ncp^{(\ip)}-1}\sum_{b=0}^{\Ncp^{(\jp)}-1} m_{a,b}^{(\ip,\jp)} \vP_a^{(\ip)}\cdot{\vP}_b^{(\jp)} - \overline{M}^{(\ip,\jp)}  \right) \left( \sum_{m=0}^{\Ncp^{(\ip)}-1} \vd^{(\ip,\jp)}_m \cdot\delta\vP^{(\ip)}_m + \sum_{n=0}^{\Ncp^{(\jp)}-1}\vd^{(\jp,\ip)}_n\cdot\delta\vP^{(\jp)}_n \right) + \Oee,
    \label{eq:tilde Jgen with R}
  \end{align}
  where
  \begin{align*}
    \vd^{(\alpha,\beta)}_m
    &:=\frac{\mu}{4\pi}\int_{C^{(\alpha)}}\diff t\int_{C^{(\beta)}}\diff t'\Biggl[ \frac{\dot{R}^{p^{(\alpha)}}_m(t)}{|\vs(t)-\vs'(t')|}\dot{\vs}'(t') - \frac{R^{p^{(\alpha)}}_m(t)\dot{\vs}(t)\cdot\dot{\vs}'(t')}{|\vs(t)-\vs'(t')|^3}(\vs(t)-\vs'(t')) \Biggr] \nonumber\\
    &=\frac{\mu}{4\pi}\sum_{k=m}^{(m+p^{(\alpha)})\% N^{(\alpha)}}\int_{I_k}\diff t\int_{C^{(\beta)}}\diff t'\Biggl[ \frac{\dot{R}^{p^{(\alpha)}}_m(t)}{|\vs(t)-\vs'(t')|}\dot{\vs}'(t') - \frac{R^{p^{(\alpha)}}_m(t)\dot{\vs}(t)\cdot\dot{\vs}'(t')}{|\vs(t)-\vs'(t')|^3}(\vs(t)-\vs'(t')) \Biggr]
  \end{align*}
  for any pair of $C^{(\alpha)}$ and $C^{(\beta)}$ such as $\alpha\ne\beta$. Further, (\ref{eq:tilde Jgen with R}) can be rearranged by the coils instead of the pairs as follows:
  \begin{align}
    \tilde{J} = J + \sum_{\alpha=1}^{\Ncoil}\sum_{m=0}^{\Ncp^{(\alpha)}-1} \bm{G}^{(\alpha)}_m \cdot \delta\vP^{(\alpha)}_m,
    \label{eq:tilde Jgen with R 2}
  \end{align}
  where the vector $\bm{G}^{(\alpha)}_m$ is the sum of all the coefficients associated with $\delta\vP^{(\alpha)}_m$.
\end{remark}

\subsection{Constraints}\label{s:constraints}

The following two types of constraints are considered in this study.

\subsubsection{Bound constraints}

\def\vPI{\overline{\vP}}
\def\DL{\bm{L}}
\def\DU{\bm{U}}

Each CP, which is a design variable, is allowed to move in a certain amount from its initial position. Regarding a CP $\vP_m$ of $C$, the bound constraint is expressed by
\begin{align}
  \DL_m \le \vP_m - \vPI_m \le \DU_m \quad\text{for $m=0$, $\ldots$, $N-1$},
  \label{eq:bounds}
\end{align}
where three-dimensional vectors $\DL_m$ and $\DU_m$ denote the lower and upper bounds of $\vP_m$, respectively, from its initial position denoted by $\vPI_m$. Here, the comparison operator '$\le$' between two vectors is applied to all their three components.

\subsubsection{Inequality constraints}

\def\length{\ell}
\def\lengthL{\length_{\rm lower}}
\def\lengthU{\length_{\rm upper}}
\def\gL{g_{\rm lower}}
\def\gU{g_{\rm upper}}

A practical constraint is to limit the length of a closed curve. For a closed curve $C$, its length, denoted by $\length$, can be calculated as 
\begin{align}
  \length
  &:=\int_C|\diff\vs| \nonumber\\
  &=\int_C\sqrt{\dot{\vs}(t)\cdot\dot{\vs}(t)}\ \diff t\nonumber\\
  &= \int_C\sqrt{\left(\sum_{m=0}^{N-1}\dot{R}^p_m(t)\vP_m\right)\cdot\left(\sum_{n=0}^{N-1}\dot{R}^p_n(t)\vP_n\right)}\ \diff t \qquad\reason{Eq.~(\ref{eq:s with R})}\nonumber\\
  &= \int_C\sqrt{\sum_{m=0}^{N-1}\sum_{n=0}^{N-1}\dot{R}^p_m(t)\dot{R}^p_n(t)\vP_m\cdot\vP_n}\ \diff t \nonumber\\
  &= \int_C\sqrt{\sum_{m=0}^{N-1}\sum_{n=0}^{N-1}\dot{R}^p_m(t)\dot{R}^p_n(t)\sum_{i=1}^3P_{m,i}P_{n,i}}\ \diff t,
  \label{eq:length-def}
\end{align}
where $\vP_m$ is expressed as $(P_{m,1},P_{m,2},P_{m,3})$ in the Cartesian coordinate system. Then, the constraint for the length of $C$ is formulated as (nonlinear) inequality constraints as follows:
\begin{align}
  \lengthL \le \length \le \lengthU
  \quad\Longleftrightarrow\quad
  \gL:= - \length + \lengthL \le 0,\quad
  \gU:= \length - \lengthU \le 0,
  \label{eq:inequality}
\end{align}
where predefined constants $\lengthL$ and $\lengthU$ denote the lower and upper lengths of $C$, respectively. When using an NLP solver (see the next section for details), it is necessary to compute the gradient of the constraint functions $\gU$ and $\gL$ with respect to the design variables, that is, the associated CPs. To this end, the differentiation of $\length$ with respect to $P_{k,l}$ (i.e. the $l$th component of $\vP_k$) is necessary and can be computed as follows:
\begin{align}
  \pp{\length}{P_{k,l}}
  &= \int_C \frac{1}{2\sqrt{\dot{\vs}(t)\cdot\dot{\vs}(t)}} \sum_{m=0}^{N-1}\sum_{n=0}^{N-1}\dot{R}^p_m(t)\dot{R}^p_n(t)\sum_{i=1}^3\left(\pp{P_{m,i}}{P_{k,l}}P_{n,i}+P_{m,i}\pp{P_{n,i}}{P_{k,l}}\right)\ \diff t\nonumber\\
  &= \int_C \frac{1}{2\sqrt{\dot{\vs}(t)\cdot\dot{\vs}(t)}} \sum_{m=0}^{N-1}\sum_{n=0}^{N-1}\dot{R}^p_m(t)\dot{R}^p_n(t)\sum_{i=1}^3\left(\delta_{mk}\delta_{il}P_{n,i}+P_{m,i}\delta_{nk}\delta_{il}\right)\ \diff t\nonumber\\
  &= \int_C \frac{1}{2\sqrt{\dot{\vs}(t)\cdot\dot{\vs}(t)}} \left(
  \sum_{n=0}^{N-1}\dot{R}^p_k(t)\dot{R}^p_n(t)P_{n,l}
  +\sum_{m=0}^{N-1}\dot{R}^p_m(t)\dot{R}^p_k(t)P_{m,l}\right)\ \diff t\nonumber\\
  &= \sum_{m=0}^{N-1} \int_C \frac{\dot{R}^p_m(t)\dot{R}^p_k(t)}{\sqrt{\dot{\vs}(t)\cdot\dot{\vs}(t)}}\diff t  P_{m,l}\nonumber\\
  &= \sum_{m=0}^{N-1} \sum_n \int_{I_n}\frac{\dot{R}^p_m(t)\dot{R}^p_k(t)}{\sqrt{\dot{\vs}(t)\cdot\dot{\vs}(t)}}\diff t  P_{m,l},
  \label{eq:diff length}
\end{align}
where $\delta_{ij}$ denotes the Kronecker delta. Also, the summation over $n$ considers the intersection of the two sets of indices involved in $\sum_{a=m}^{(m+p)\% N}$ and $\sum_{b=m}^{(k+p)\% N}$, which are the supports of the basis functions $R^p_m(t)$ and $R^p_k(t)$, respectively.

\subsection{Utilising a gradient-based NLP solver}\label{s:NLP}

\def\mx{\mathsf{x}}
\def\mxL{\mathsf{x}_{\rm lower}}
\def\mxU{\mathsf{x}_{\rm upper}}
\def\mD{\mathsf{D}}
\def\Ncon{N_{\rm constraint}}
\def\mG{\mathsf{G}}

The shape optimisation problem mentioned in Section~\ref{eq:mutual_problem} can be solved with a general-purpose and gradient-based NLP solver that can find an $m$-dimensional design variable, say $\mx$, so that it can minimise (or maximise) a certain objective function, say $f(\mx)$, subject to bound constraints such as $\mxL\le\mathsf{x}\le\mxU$ (where $\mxL$ and $\mxU$ are prescribed bounds; the component-wise operation was applied) and inequality constrains such as $h_i(\mx)\le 0$ for $i=0$, $1$, $\ldots$.

To utilise such an NLP solver for minimising the general-case $J$ in (\ref{eq:Jgen}), the design variable vector $\mx$ should be treated as the compound of all the CPs in all the $\Ncoil$ coils, i.e. $\vP_i^{(\alpha)}$, where $i=0,\ldots,\Ncp^{(\alpha)}-1$ for every $\alpha=1$, $\ldots$, $\Ncoil$. Specifically, $\mx$ can be expressed as
\begin{align*}
  \mathsf{x}
  :=\left(\mx^{(1)},\ \cdots,\ \mx^{(\Ncoil)}\right)\in\mathbb{R}^{3m},
\end{align*}
where the sub-vector $\mx^{(\alpha)}$ consists of the CPs of $C^{(\alpha)}$, i.e.
\begin{align*}
  \mx^{(\alpha)}
  :=\left(\vP^{(\alpha)}_0,\ldots,\vP^{(\alpha)}_{\Ncp^{(\alpha)}-1}\right)
  \equiv\left(P^{(\alpha)}_{0,1}, P^{(\alpha)}_{0,2}, P^{(\alpha)}_{0,3}, \cdots,P^{(\alpha)}_{N^{(\alpha)}-1,1}, P^{(\alpha)}_{N^{(\alpha)}-1,2}, P^{(\alpha)}_{N^{(\alpha)}-1,3}\right)\in\mathbb{R}^{3\Ncp^{(\alpha)}}.
\end{align*}
Also, the number $m$ denotes the total number of the CPs, i.e. $m:=\sum_{\alpha=1}^{\Ncoil}\Ncp^{(\alpha)}$.

Then, $\tilde{J}$ in (\ref{eq:tilde Jgen with R 2}) can be written as
\begin{align*}
  \tilde{J}(\mx) = J(\mx) + \nabla J(\mx)\cdot\delta\mx,
\end{align*}
where 
\begin{align*}
  \nabla J(\mx) := \left(\mG^{(1)},\ldots,\mG^{(\Ncoil)} \right)\in\mathbb{R}^{3m}
\end{align*}
denotes the gradient of $J$ with respect to $\mx$ and the $k$th sub-vector consists of the sensitivities for the CPs of $C^{(\alpha)}$, i.e.
\begin{align*}
  \mG^{(\alpha)}:=\left(\bm{G}_0^{(\alpha)},\ldots,\bm{G}_{\Ncp^{(\alpha)}-1}^{(\alpha)}\right)\in\mathbb{R}^{3\Ncp^{(\alpha)}}.
\end{align*}

In addition, the inequality constraints in (\ref{eq:inequality}) can be represented with using $\mx$ as follows:
\begin{align}
  \gL^{(\alpha)}(\mx) &:= - \length^{(\alpha)}(\mx) + f_{\rm lower}^{(\alpha)}\length^{(\alpha)}(\mx_{\rm init}) \le 0,\nonumber\\
  \gU^{(\alpha)}(\mx) &:=  \length^{(\alpha)}(\mx) - f_{\rm upper}^{(\alpha)}\length^{(\alpha)}(\mx_{\rm init}) \le 0
  \label{eq:length}
\end{align}
for any $\alpha\in[1,\Ncoil]$, where $\length^{(\alpha)}$ is computed from (\ref{eq:length-def}) where $C$ is considered as $C^{(\alpha)}$. Also, the predefined constants $f_{\rm lower}^{(\alpha)}$ and $f_{\rm upper}^{(\alpha)}$ determines the lower and upper bounds of the length of $C^{(\alpha)}$ relatively to the initial length $\length^{(\alpha)}(\mx_{\rm init})$, respectively. The gradient of $\gL^{(\alpha)}$ with respect to $\mx$ can be represented as $3m$-dimensional vector as follows:
\begin{align*}
  \nabla \gL^{(\alpha)}(\mx)
  &:=\Biggl(
  \underbrace{0,\ldots,0,}_{\text{$C^{(1)}$, $\ldots$, $C^{(\alpha-1)}$}}
  \ \underbrace{\pp{\gL^{(\alpha)}}{P^{(\alpha)}_{0,1}}, \pp{\gL^{(\alpha)}}{P^{(\alpha)}_{0,2}}, \pp{\gL^{(\alpha)}}{P^{(\alpha)}_{0,3}}, \ldots,\pp{\gL^{(\alpha)}}{P^{(\alpha)}_{N^{(\alpha)}-1,1}}, \pp{\gL^{(\alpha)}}{P^{(\alpha)}_{N^{(\alpha)}-1,2}}, \pp{\gL^{(\alpha)}}{P^{(\alpha)}_{N^{(\alpha)}-1,3}},}_{\text{Derivatives for the CPs of $C^{(\alpha)}$}}
  \ \underbrace{0,\ldots,0}_{\text{$C^{(\alpha+1)}$, $\ldots$, $C^{(\Ncoil)}$}}
  \Biggr)\nonumber\\
  &=\Biggl(
  0,\ldots,0,
  \quad -\pp{\length^{(\alpha)}}{P^{(\alpha)}_{0,1}}, -\pp{\length^{(\alpha)}}{P^{(\alpha)}_{0,2}}, -\pp{\length^{(\alpha)}}{P^{(\alpha)}_{0,3}}, \ldots,-\pp{\length^{(\alpha)}}{P^{(\alpha)}_{N^{(\alpha)}-1,1}}, -\pp{\length^{(\alpha)}}{P^{(\alpha)}_{N^{(\alpha)}-1,2}}, -\pp{\length^{(\alpha)}}{P^{(\alpha)}_{N^{(\alpha)}-1,3}},
  \quad 0,\ldots,0
  \Biggr)
\end{align*}
where the differentiation of $\length^{(\alpha)}$ with respect to $P^{(\alpha)}_{k,l}$ can be computed with (\ref{eq:diff length}) where $C$ and $P$ are regarded as $C^{(\alpha)}$ and $P^{(\alpha)}$, respectively. The gradient of $\gU^{(\alpha)}$ is expressed similarly.

The solution of $\mx$ can be searched by a gradient-based NLP solver whenever the user provides the routines to compute $J$, $\nabla J$, $\gL^{(\alpha)}$, $\gU^{(\alpha)}$, $\nabla \gL^{(\alpha)}$, and $\nabla \gU^{(\alpha)}$, where $\alpha$ considers all the coils to which the inequality constraints are applied.

This study utilises a gradient-based NLP named SLSQP (sequential least-squares quadratic programming) in an optimisation library NLopt~\cite{nlopt}. The library was incorporated to our computer program written in the C language.

\section{Numerical examples}\label{s:num}

This section assesses the proposed SO system through some examples. In all the examples, without the loss of generality, the ambient permeability $\mu$ is assumed to be $1$ \si{\henry/\meter}; SI units will be omitted for numerical values hereafter. This assumption means that the magnetic field $\vB$ in (\ref{eq:B}), the MI $M$ in (\ref{eq:M}), and the shape derivative $D$ (of $M$) in (\ref{eq:D}) are normalised by $\mu$, while the objective function $J$ in (\ref{eq:J}) or (\ref{eq:Jgen}) and its shape derivative $G$ in (\ref{eq:G}) are normalised by $\mu^2$.

\subsection{Example 1: Verification of the shape derivative $D$ in (\ref{eq:D}). }

The first example verifies the proposed SO framework by considering the SO whose theoretical solution can be computed.

\subsubsection{Problem setting}

\begin{figure}[hbt]
  \centering
  \includegraphics[width=.7\textwidth]{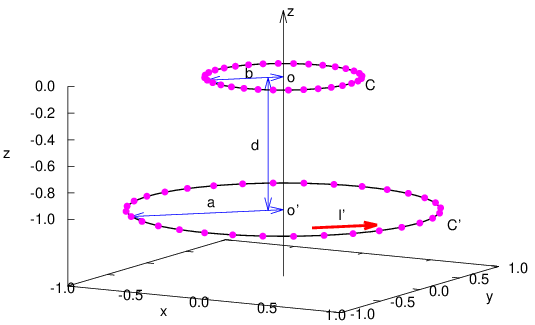}
  \caption{Example 1: Problem setting to verify the shape sensitivity $\frac{\diff M}{\diff b}$ for the MI $M$ associated with the two coaxial coils $C$ and $C'$ with the radii $b$ and $a$, respectively, at the distance of $d$. This figure corresponds to the case of $a=1.0$, $b=0.5$, and $d=1.0$. The $32\times 2$ points represent the CPs in the case of $N=N'=32$.}
  \label{fig:verification2-config}
\end{figure}

As shown in Figure \ref{fig:verification2-config}, let us suppose two coaxial circular coils $C$ (receiver) and $C'$ (transmitter) on the plane $z=0$ and $z=-d$, respectively, where $d$ ($>0$) stands for the distance between two coils in the $z$-direction.The centre and radius of $C$ are the origin $\bm{o}:=(0,0,0)$ and $b$ ($>0$), respectively, while those of $C'$ are $\bm{o}':=(0,0,-d)$ and $a$ ($>0$), respectively. In addition, the steady current $I'$ of the magnitude $1.0$ is running on $C'$ in the counter clock-wise when viewed from the positive $z$-axis. Letting this configuration as the initial state, the present SO designs the radius $b$ of $C$ so that the following objective function $J$ is \textit{maximised}:
\begin{align}
  J = \frac{M^2}{2},
\end{align}
where $M$ is the MI of the two coils and $J$ corresponds to the case of $\overline{M}=0$ in (\ref{eq:J}). 

Then, the derivative of $M$ with respect to $b$, i.e. $\frac{\diff M}{\diff b}$, can be computed analytically for any $b$. Meanwhile, it can be numerically evaluated by using the shape sensitivities $\vd$ and $\vd'$ in (\ref{eq:vd}). Hence, if their results for $\frac{\diff M}{\diff b}$ agree with each other, the shape sensitivities and thus the shape derivative $D$ in (\ref{eq:D}), which is the core of the proposed SO framework, can be verified.

First, the sensitivity $\frac{\diff M}{\diff b}$ is calculated analytically. In terms of the cylindrical coordinate system $O-\rho\theta z$, the vector potential $\vA=\vA(\rho,\theta,z)$ is given in \cite{jackson1999classical}, for example, as
\begin{align}
  A_\rho(\rho,\theta,z) = A_z(\rho,\theta,z) = 0,\qquad
  A_\theta(\rho,\theta,z) = \frac{\mu I'}{\pi k}\sqrt{\frac{a}{\rho}}\left[\left(1-\frac{k^2}{2}\right)K(k^2)-E(k^2)\right],
\end{align}
where $k^2:=\frac{4a\rho}{(a+\rho)^2+z^2}$ and $K$ and $E$ are the complete elliptic integrals of the first and second kinds~\cite{abramowitz1972}, respectively, that is,
\begin{align*}
  K(m):=\int_0^{\pi/2}\frac{\diff\phi}{\sqrt{1-m\sin^2\phi}},\qquad
  E(m):=\int_0^{\pi/2}\sqrt{1-m\sin^2\phi}\ \diff\phi.
\end{align*}
Therefore, from (\ref{eq:Phi}) and (\ref{eq:M}), the MI between $C$ and $C'$ can be calculated exactly as
\begin{align*}
  M(b)
  &=\frac{1}{I'}\int_C\vA(b,\theta,d) \cdot\diff\vs
  =\frac{1}{I'}\int_0^{2\pi}A_\theta(b,\theta,d) b\diff\theta
  =\frac{2\mu b}{k}\sqrt{\frac{a}{b}}\left[\left(1-\frac{k^2}{2}\right)K(k^2)-E(k^2)\right],
\end{align*} 
where $k^2=\frac{4ab}{(a+b)^2+d^2}$ because $\rho=b$ and $z=d$ hold on $C$. Then, the sensitivity of $M$ with respect to the radius $b$ is calculated as
\begin{align}
  \frac{\diff M}{\diff b}
  =\frac{\mu k}{2}\sqrt{\frac{b}{a}}\left[\frac{a^2-b^2-d^2}{(a-b)^2+d^2}E(k^2)+K(k^2)\right].
  \label{eq:dMdb_exact}
\end{align}

Meanwhile, in order to compute $\frac{\diff M}{\diff b}$ within the proposed framework, it is necessary to relate the location of each CP to the radius $b$ at any iteration step. Suppose the CPs on $C$ and $C'$ are on the circle with the radii $b$ and $a$, respectively.\footnote{
Strictly speaking, the resulting coils $C$ and $C'$ deviate from perfect circles of radii $b$ and $a$, respectively, due to the inherent characteristics of closed B-spline curves. However, this deviation becomes negligible with an increasing number of control points.} Then, as $b$ is perturbed to $\tilde{b}$, the $m$th CP $\vP_m$ of $C$ needs to change proportionally, that is,
\begin{align*}
  \tilde{\vP}_m = \frac{\tilde{b}}{b}(\vP_m-\bm{o})+\bm{o}.
\end{align*}
Therefore, the perturbation of $\vP_m$ can be expressed as
\begin{align*}
  \delta\vP_m := \tilde{\vP}_m - \vP_m = \frac{\tilde{b}-b}{b}(\vP_m-\bm{o}) =\frac{\delta b}{b}(\vP_m-\bm{o}),
\end{align*}
where $\delta b:=\tilde{b}-b$. Since the radius $a$ of $C'$ is fixed, $\delta\vP'_n$ is always zero for any $n$. Therefore, (\ref{eq:tilde M with R}) is expressed as follows:
\begin{align*}
  \tilde{M}
  &=
  M
  +\left(\sum_{m=0}^{N-1}\frac{\vd_m\cdot(\vP_m-\bm{o})}{b}\right)\delta b
  +0
  +\Oee.
\end{align*}
In the RHS, the coefficient of $\delta b$ corresponds to the sensitivity of $M$ with respect to $b$, that is,
\begin{align}
  \frac{\diff M}{\diff b}
  =\sum_{m=0}^{N-1}\frac{\vd_m\cdot(\vP_m-\bm{o})}{b}
  \label{eq:dMdb_num}
\end{align}
Successively, this is compared with (\ref{eq:dMdb_exact}) to check the correctness of the sensitivity $\vd_m$ derived in (\ref{eq:vd}).

\subsubsection{Numerical results}

The exact sensitivity $\frac{\diff M}{\diff b}$ in (\ref{eq:dMdb_exact}) and the numerical counter part in (\ref{eq:dMdb_num}) are compared for various $b$ when $a=1.0$, $d=1.0$, and the degree of the B-spline functions is two, i.e. $p=p'=2$. In this paper, changes in degree were not investigated, and no numerical implementation was performed for degrees higher than $2$; thus, degree of $2$ is used in all the examples. It is likely that the results would remain consistent for higher degrees, particularly for degree of $3$, provided that the number of CPs is not exceptionally small. Further, the number, $Q$, of Gauss--Legendre quadrature points per interval is fixed to $16$, which will be used through this paper. The numerical justification for the sufficiency of $Q=16$ stems from the observation that the results for $Q=16$, $8$, $4$ were nearly identical in this specific example. Neverthless, there is no theoretical guarantee that $Q=16$ will always be sufficient for other configurations, as the relevant integral kernels are rational functions, not polynomials. Although computationally more expensive, it would be preferable to utilise an adaptive integral, such as the Gauss--Kronrod quadrature formula; however, this was not implemented.

\begin{figure}[h]
  \centering
  \includegraphics[width=.6\textwidth]{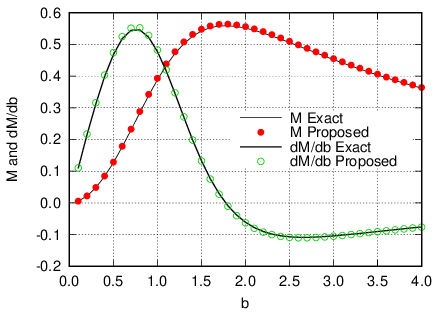}
  \caption{Example 1: Comparison of the sensitivity $\frac{\diff M}{\diff b}$ as well as the MI $M$.}
  \label{fig:verification2-ng16ncp32}
\end{figure}

Figure~\ref{fig:verification2-ng16ncp32} shows the sensitivity $\frac{\diff M}{\diff b}$ for $b\in[0.1, 4.0]$ with using $32$ CPs per coil, i.e. $N=N'=32$. A good agreement is observed, which is true also for $M$ as in the same figure. 

\begin{figure}[h]
  \centering
  \includegraphics[width=.6\textwidth]{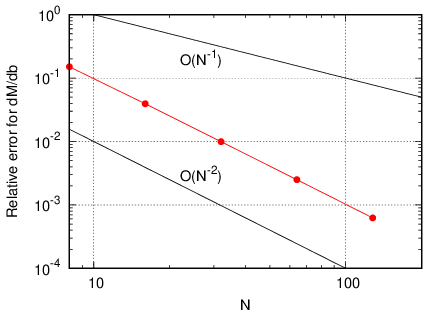}
  \caption{Example 1: Comparison of the relative error of the sensitivity $\frac{\diff M}{\diff b}$ in terms of the number $N$ ($\equiv N'$) of CPs when the radius $b$ is fixed to $1.0$.}
  \label{fig:verification2-ng16}
\end{figure}

In addition, Figure~\ref{fig:verification2-ng16} plots the relative error of the numerical sensitivity to the exact one against $N$, with supposing $N'\equiv N$ and fixing $b$ to $1.0$. It is confirmed that the relative error decreases as $N$ increases; the rate is approximately $O(N^{-2})$. These results can verify the shape sensitivities in (\ref{eq:vd}) and thus the shape derivative $D$ in (\ref{eq:D}).

In general, there is no specific way to definitively know if the number $N$ of CPs is sufficient. For this parametric SO, which uses the radius $b$ as a design variable, one could potentially leverage the geometrical deviation between a closed curve and the true circle determined by a candidate $b$. Generally, however, the number (and positions) of CPs must be determined so that the resulting closed curve can adequately represent both a given initial shape and an unknown optimised shape.

\nprounddigits{3} 

Next, the present maximisation problem is solved in terms of the solo design variable $b$. The optimal value of $b$, say $b^*$, and the corresponding MI $M^*$ are read as $\numprint{1.770000e+00}$ and $\numprint{5.640263e-01}$, respectively, from Figure~\ref{fig:verification2-ng16ncp32}. Therefore, the optimal value of $J$ is $J^*:=\frac{(M^*)^2}{2}=\numprint{1.5906283354584500000e-01}$.

The SLSQP solver (recall Section~\ref{s:NLP}) is terminated when the relative change of $J$ becomes less than $10^{-5}$ or when the number of optimisation steps exceeds $1000$, although the latter did not occur actually. The initial value of $b$ is selected as either $1.0$ and $3.0$. Also, the number $N$ ($\equiv N'$) of CPs in each coil is set to $32$ or $64$.

\begin{figure}[h]
  \centering
  \includegraphics[width=.6\textwidth]{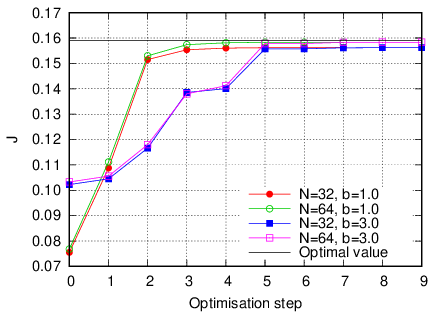}
  \caption{Example 1: History of $J$ for the four cases. The black line presents the optimal value $J^*$.}
  \label{fig:work45-J}
\end{figure}

Figure~\ref{fig:work45-J} shows the history of $J$. For every $N$, the optimal state was obtained regardless of the initial values. It is observed that the results of $N=64$ are slightly more accurate than $N=32$, which is reasonable from the viewpoint of the discretisation error. Table~\ref{tab:work45} summarises the present optimisation.

\begin{table}[h]
  \nprounddigits{3}
  \centering
  \caption{Example 1: The optimised $b$ and $J$.}
  \begin{tabular}{cccc}
    \hline
    \# of CPs $N$ & Initial $b$ & Optimised $b$ & Optimised $J$\\
    \hline
    32 & 1.0 & \numprint{1.775715e+00} & \numprint{1.562018e-01}\\
    64 & 1.0 & \numprint{1.771563e+00} & \numprint{1.583430e-01}\\
    32 & 3.0 & \numprint{1.775765e+00} & \numprint{1.562018e-01}\\
    64 & 3.0 & \numprint{1.771625e+00} & \numprint{1.583430e-01}\\
    \hline
    Optimal & & \numprint{1.770000e+00} & \numprint{1.5906283354584500000e-01}\\
    \hline
  \end{tabular}
  \label{tab:work45}
\end{table}

\subsection{Example 2: Shape optimisation with using inequality constraints for length}\label{s:work41}

This example considers the inequality constrains with regard to the coil length, which was given by (\ref{eq:length}).

\subsubsection{Problem setting}

At the initial step, a circular coils $C$ (respectively, $C'$) with radius of $2.0$ (respectively, $1.0$) and centre of $(1.0, 0.0, 1.0)$ (respectively, $(0.0, 0.0, 0.0)$) is placed on the plane $z=1.0$ (respectively, $z=0.0$). Then, only $C$ is optimised so that the mutual inductance $M$ between $C$ and $C'$ is $0.1$ ($=\overline{M}$). Each CP of $C$ can move freely in the $x$ and $y$ directions but the vertical movement is restricted to $\pm 0.5$ from its initial position. Further, the length of $C$ must keep its initial length within $1$\%; this means $f^{(0)}_{\rm lower}=0.99$ and $f^{(0)}_{\rm lower}=1.01$ in (\ref{eq:length}), where the superscript `$(0)$' corresponds to $C$. Specifically, the initial length is \numprint{12.50594}; it is slightly smaller than $4\pi$ ($=\numprint{12.56637061435917295376}$) because the CPs are placed on the circle with radius of $2.0$ and thus the radius of the resulting and approximated circle is smaller than $2.0$.

Analogous to Example 1, each coil is modelled by the B-spline closed curve of degree $2$ using $32$ CPs, i.e. $p=p'=2$ and $N=N'=32$ are employed.
Furthermore,  the stopping criterion of the SLSQP solver remains identical to that in Example 1.

\subsubsection{Results and discussions}

\begin{figure}[hbt]
  \centering
  \includegraphics[width=.6\textwidth]{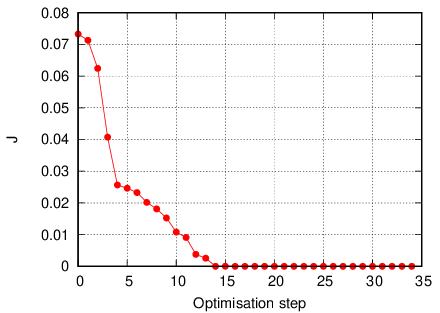}
  \caption{Example 2: History of $J$. The final value of $J$ is \numprint{3.466674e-33}.}
  \label{fig:work41-J}
\end{figure}

\begin{figure}[hbt]
  \centering
  \includegraphics[width=.6\textwidth]{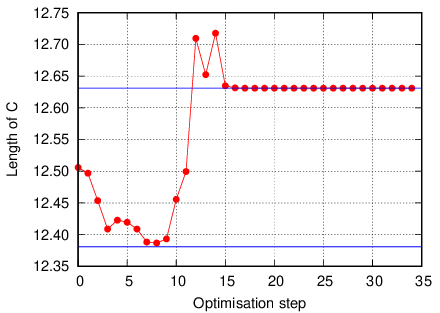}
  \caption{Example 2: History of the length of $C$. The two horizontal lines in blue are the given bounds, i.e. $\numprint{12.3808806}$ and $\numprint{12.6309994}$.}
  \label{fig:work41-length}
\end{figure}

\begin{figure}[hbt]
  \centering
  \includegraphics[width=.8\textwidth]{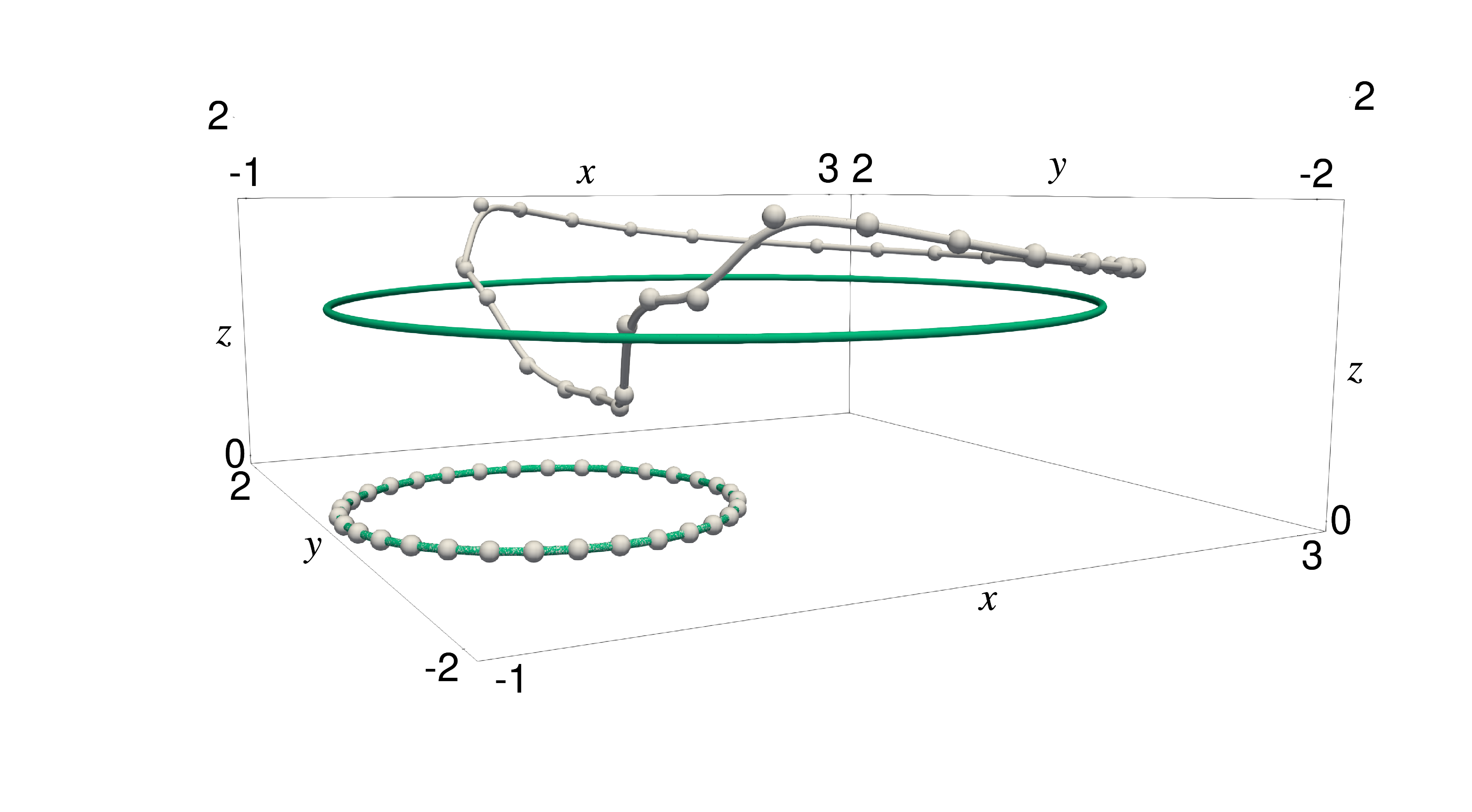}
  \caption{Example 2: Initial shape (green) and optimised shape (silver). The points represent the CPs of the optimised shape. The lower coil $C'$ on $z=0$ is not optimised in the present optimisation. For the reference, a bounding box of $[-1, 3]\otimes[-2, 2]\otimes[0, 2]$ is shown.}
  \label{fig:work41-shape-init-final}
\end{figure}

\begin{figure}[hbt]
  \centering
  \begin{tabular}{cc}
    Initial shape & Optimised shape\\
    \includegraphics[width=.45\textwidth]{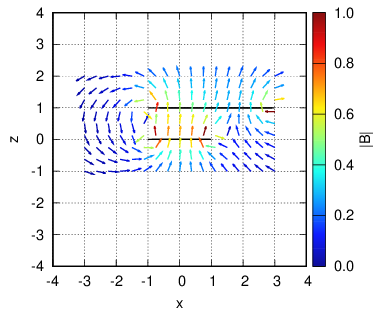}
    &\includegraphics[width=.45\textwidth]{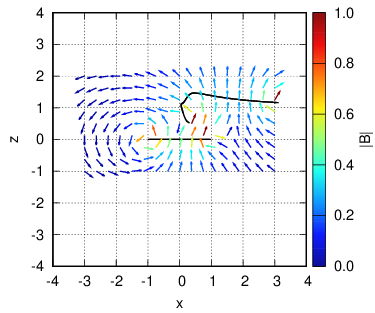}
  \end{tabular}
  \caption{Example 2: The magnetic field $\vB$ on the $xz$-plane (specifically, the region such as $-3.0\le x\le 3.0$, $y=0.0$, and $-1.0\le z \le 2.0$) for the initial and optimised shapes. The black lines are the projection of the two coils onto the $xz$-plane. The colour in the vectors presents the strength $|\vB|$, which was truncated to $1.0$ if the value exceeded $1.0$.} 
  \label{fig:work41-B}
\end{figure}

The iteration converged successfully at the $34$th step, as shown in Figure~\ref{fig:work41-J}. The history of the length of $C$ in Figure~\ref{fig:work41-length} shows that the underlying constraint is satisfied. Moreover, the optimised shape is compared with the initial one in Figure~\ref{fig:work41-shape-init-final}. Since the MI at the initial step is larger than the target value $\overline{M}$, the target coil $C$ bent to the fixed coil $C'$ in order to prevent the magnetic field from going through $C$ in the $+z$ direction. This can be observed in Figure~\ref{fig:work41-B}.

\subsection{Example 3: Toroidal coil}\label{s:toroidal}

The final example treats a more complicated SO, considering a magnetic confinement into a toroidal coil.

\subsubsection{Problem setting}

\begin{figure}[h]
  \centering 
  \includegraphics[width=.7\textwidth]{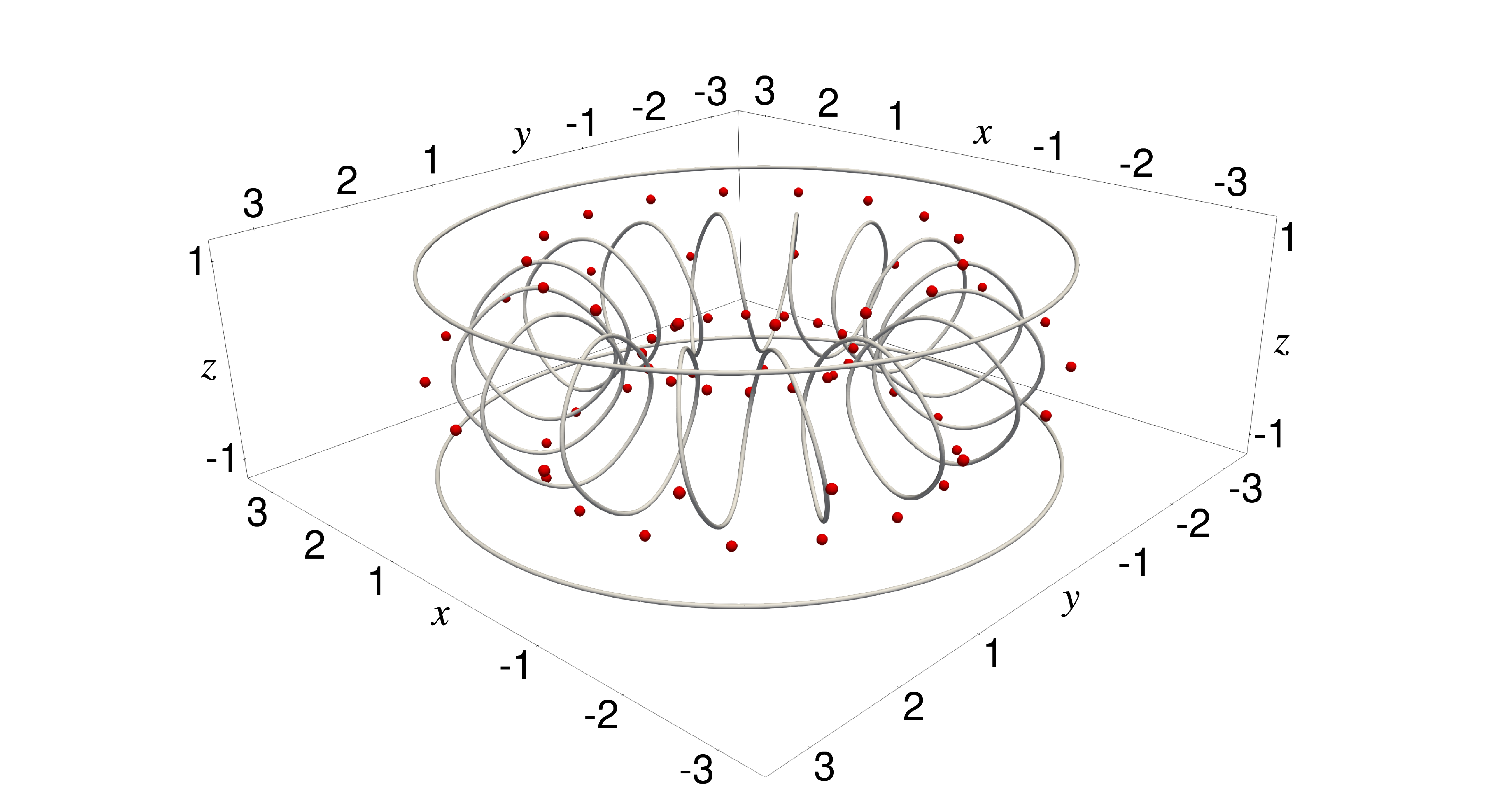}
  \caption{Example 3: Configuration. The red points are the CPs for the toroidal coil $C^{(1)}$ and designed in this optimisation. The inner radius $a$ and outer radius $b$ are $1.0$ and $2.0$ respectively. Then, the height is $2.0$ ($=2b$). The upper and lower circular coils with the radius of $3.0$ ($=a+b$) are $C^{(2)}$ and $C^{(3)}$, respectively, and are not optimised. For the reference, a bounding box of $[-3.45,3.45]\otimes[-3.45,3.45]\otimes[-1.15,1.15]$ is shown.}
  \label{fig:toroidal-config}
\end{figure}

Let us consider a system of three coils, as shown in Figure~\ref{fig:toroidal-config}. The first coil, denoted by $C^{(1)}$ in terms of the general notation introduced in Remark~\ref{remark:gen}, is a toroidal coil to be optimised and any point $\bm{p}\in C^{(1)}$ is given by
\begin{align*}
  \bm{p}(t):=\begin{bmatrix}
  \left(a-b\cos(f t)\right)\cos t\\
  \left(a-b\cos(f t)\right)\sin t\\
  b\sin(f t)
  \end{bmatrix}\quad\text{for $t\in[0,2\pi]$},
\end{align*}
where $a$ and $b$ are the outer and inner radii, respectively, and $f$ is the frequency of the coil or the number of turns. In what follows, $a=2.0$, $b=1.0$, and $f=16$ are supposed. The remaining two coils $C^{(2)}$ and $C^{(3)}$ are circular coils with radius of $a+b$ ($=3.0$) and on the planes of $z=-b$ ($=-1.0$) and $z=b$ ($=1.0$), respectively, so that the toroidal coil $C^{(1)}$ is bounded by them in the $z$-direction. Since the current is not provided to both $C^{(2)}$ and $C^{(3)}$, they do not generate any magnetic field but play a role to measure the MIs with $C^{(1)}$ due to the possible magnetic field through the ``hole'' of $C^{(1)}$ along the $z$-axis. To confine the magnetic field within $C^{(1)}$, one may minimise those MIs as much as possible. Specifically, from the general expression in (\ref{eq:Jgen}), the present objective function $J$ is written as
\begin{align}
  J = \frac{1}{2}\left\{\left(M^{(1,2)}\right)^2+\left(M^{(1,3)}\right)^2\right\},
\end{align}
where $M^{(1,2)}$ (respectively, $M^{(1,3)}$) denotes the MI between $C^{(1)}$ and $C^{(2)}$ (respectively, $C^{(3)}$). Here, the target values $\overline{M}^{(1,2)}$ and $\overline{M}^{(1,3)}$ are given as zero.

All the coils are expressed with the B-spline functions of degree $2$, i.e. $p^{(1)}=p^{(2)}=p^{(3)}=2$. The toroidal coil $C^{(1)}$ consist of $64$ CPs, i.e. $N^{(1)}=64$, while each of $C^{(2)}$ and $C^{(3)}$ consists of $32$ CPs, i.e. $N^{(2)}=N^{(3)}=32$. All the other numerical settings are the same as those in the previous examples.

For comparison, the following three cases of constraints are considered:
\begin{description}
\item[Case I] Each CP for $C^{(1)}$ can move $\pm0.2$ ($=\pm\Delta$) from its initial position in both the $x$ and $y$ directions but cannot move in the $z$ direction.
\item[Case II] The value of $\Delta$ is replaced with $0.3$ instead of $0.2$ of the previous case.
\item[Case III] The initial length of $C^{(1)}$, i.e. $\numprint{74.44167}$, can vary $\pm 0.1\%$, although any CP cannot move in the $z$ direction.
\end{description}

\subsubsection{Results and discussions}

\begin{figure}[hbt]
  \centering
  \begin{tabular}{cc}
    \includegraphics[width=.45\textwidth]{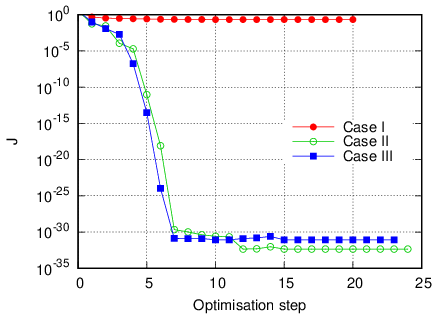}
    &\includegraphics[width=.45\textwidth]{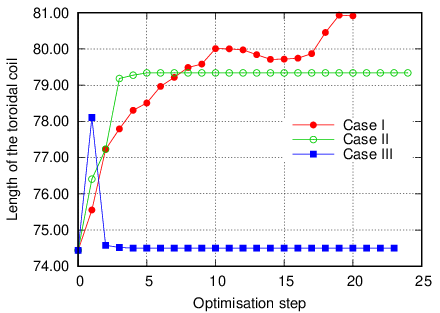}
  \end{tabular}
  \caption{Example 3: (Left) History of $J$. (Right) History of the length of the toroidal coil $C^{(1)}$.}
  \label{fig:toroidal-J}
\end{figure}

As shown in Figure~\ref{fig:toroidal-J} (left), the objective function $J$ did not become small in Case I, while it became sufficiently small in Cases II and III. This means that the variation $\Delta=0.2$ for Case I was insufficient to minimise $J$. As for Case III, Figure~\ref{fig:toroidal-J} (right) shows that the length of $C^{(1)}$ increased \numprint{0.083394152764170927}\% from its initial value, i.e. \numprint{74.44167}.

\begin{figure}[p]
  \centering
  \begin{tabular}{p{30pt}cc}
    & Bird view & Top view\\
    Initial shape
    &\includegraphics[width=.4\textwidth, valign=t]{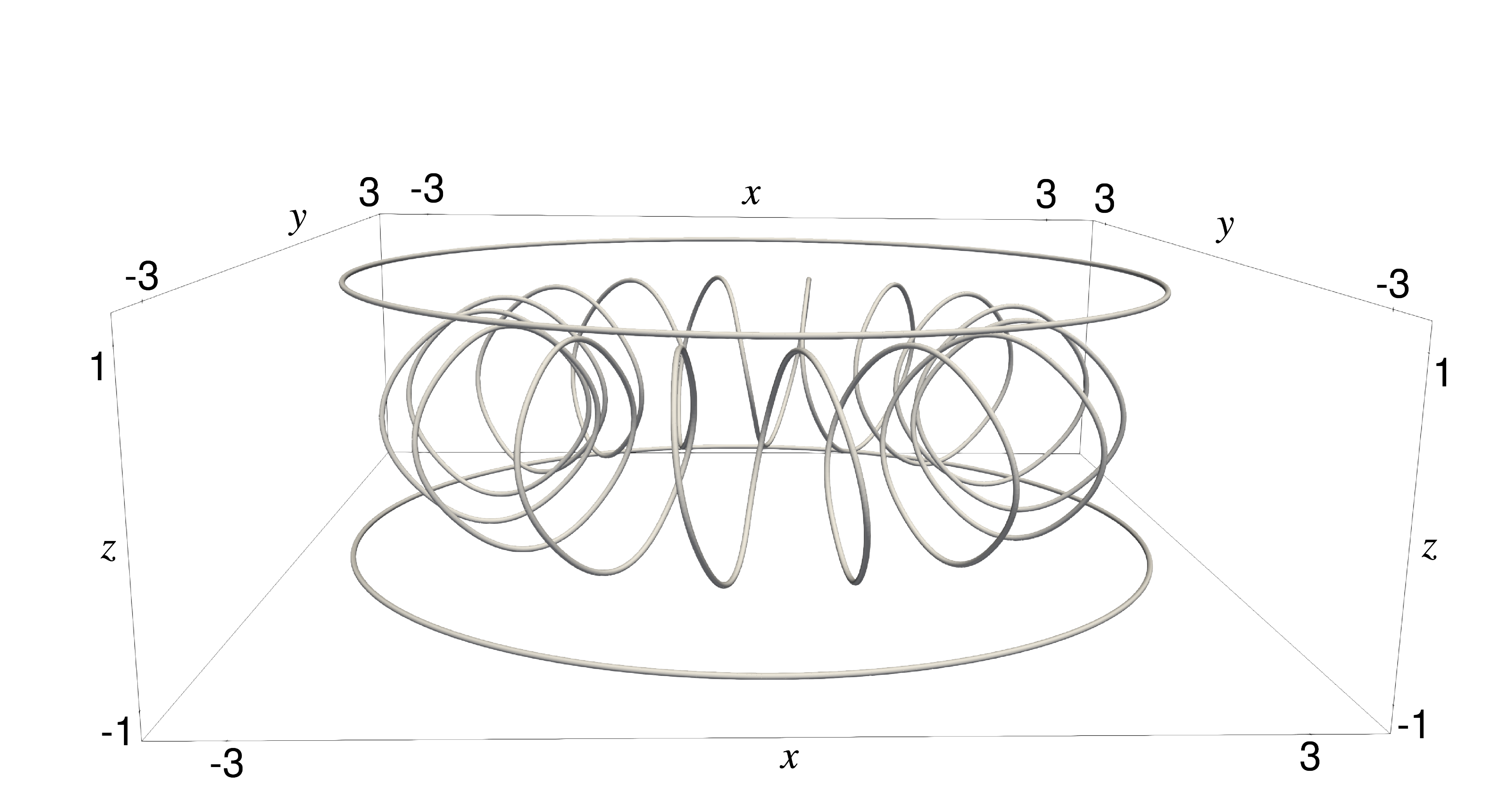}
    &\includegraphics[width=.5\textwidth, valign=t]{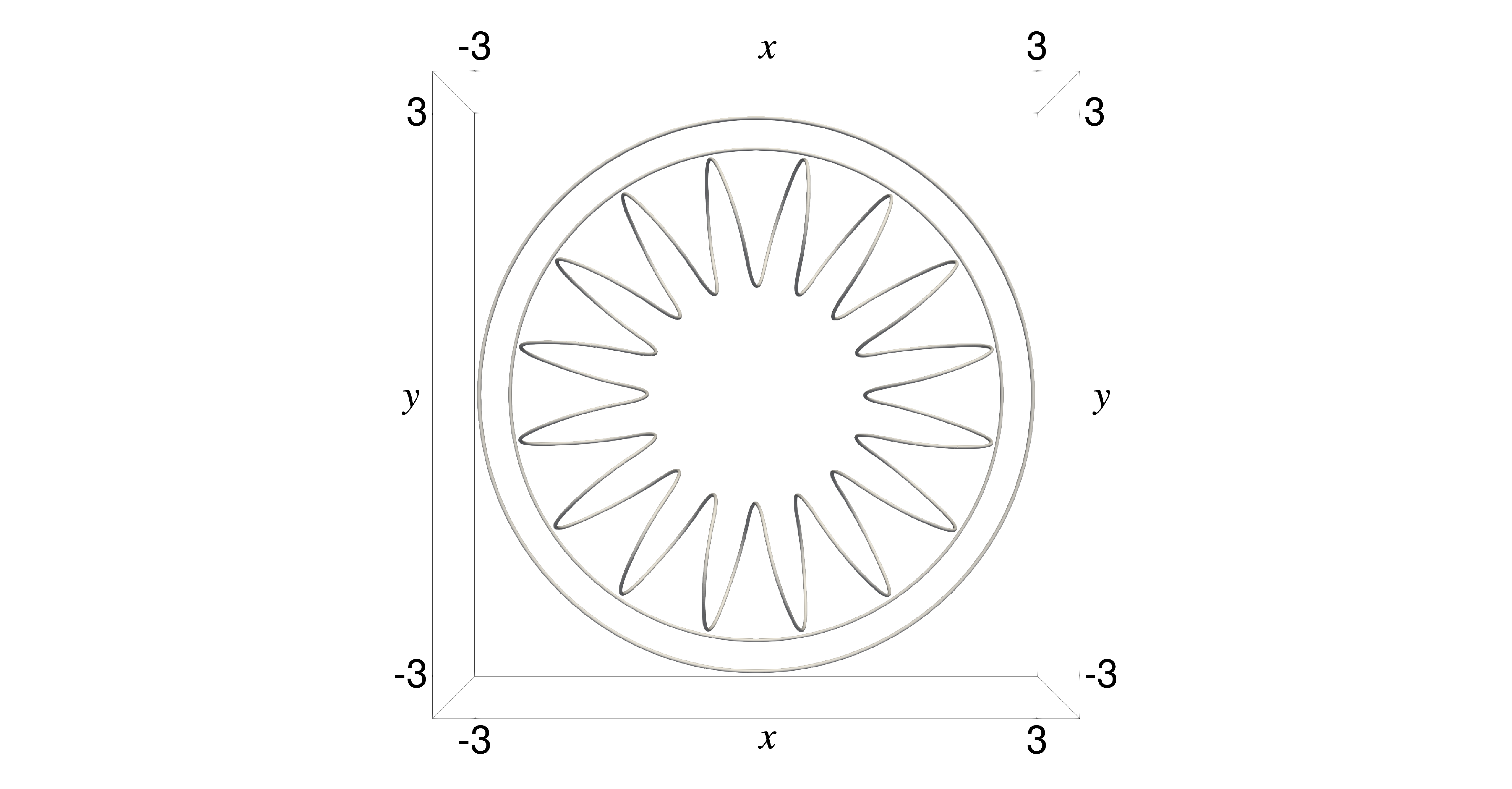}\\ 
    Optimised shape\newline\mbox{(Case I)}
    &\includegraphics[width=.4\textwidth, valign=t]{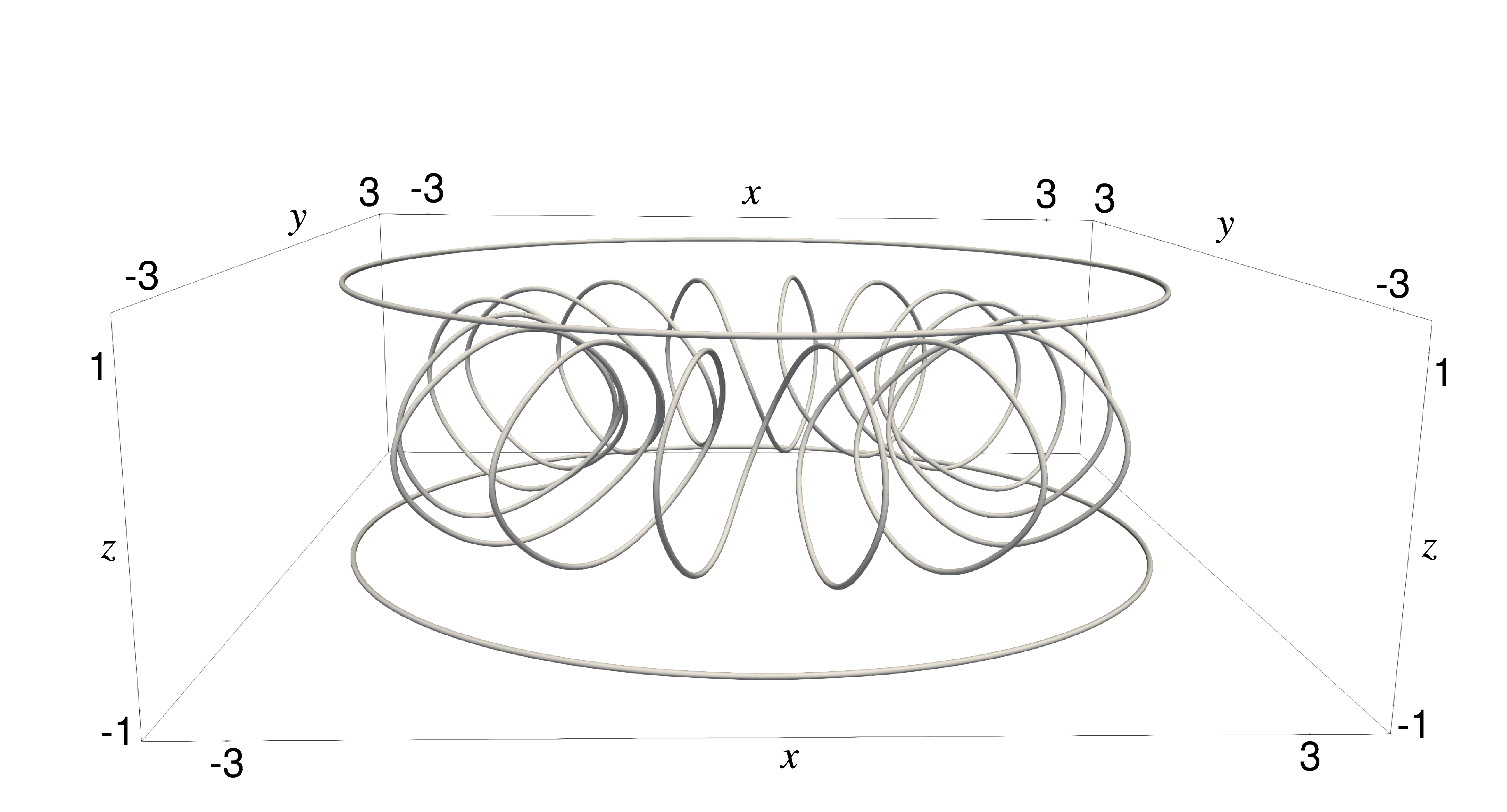}
    &\includegraphics[width=.5\textwidth, valign=t]{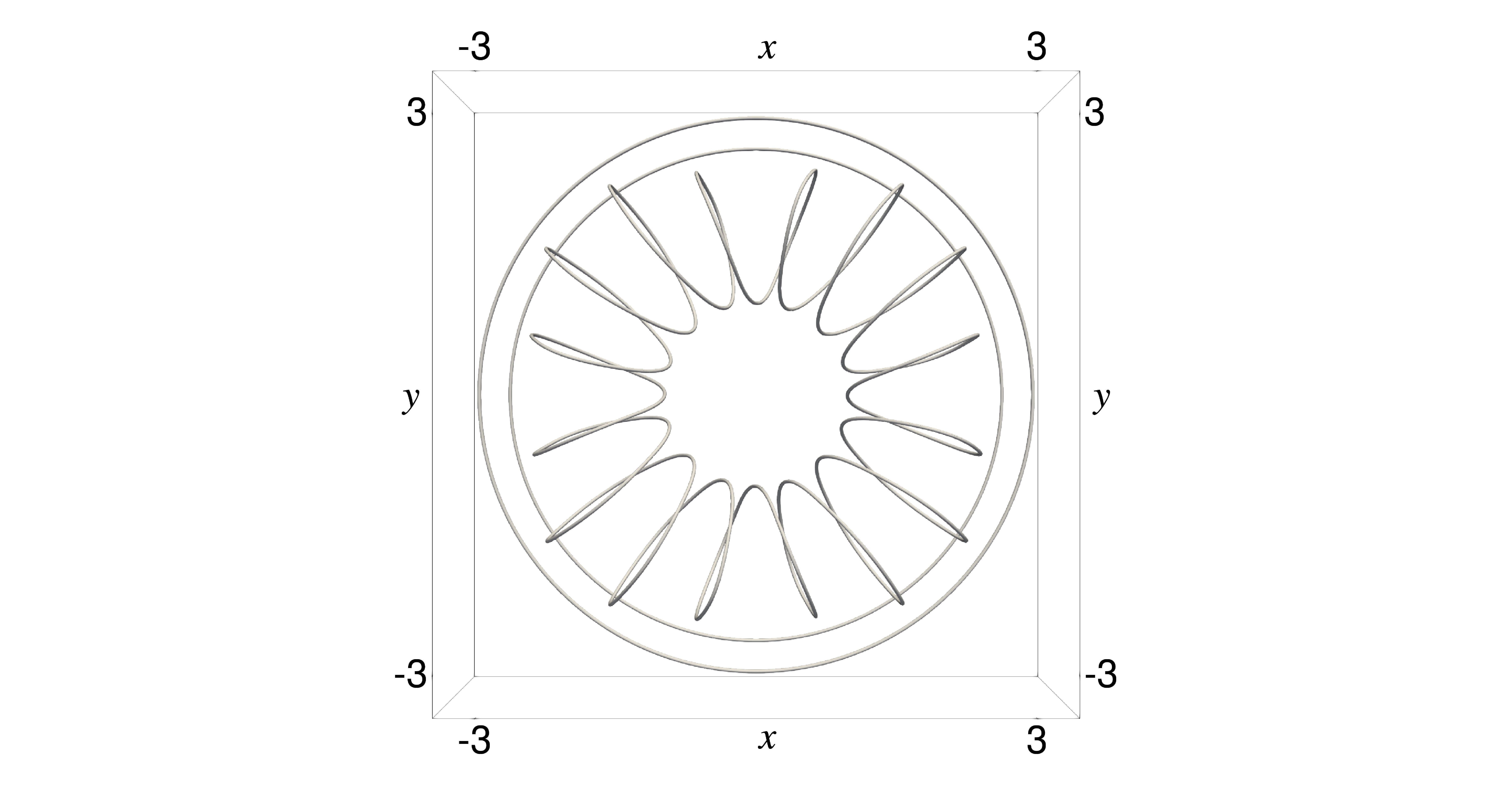}\\ 
    Optimised shape\newline\mbox{(Case II)}
    &\includegraphics[width=.4\textwidth, valign=t]{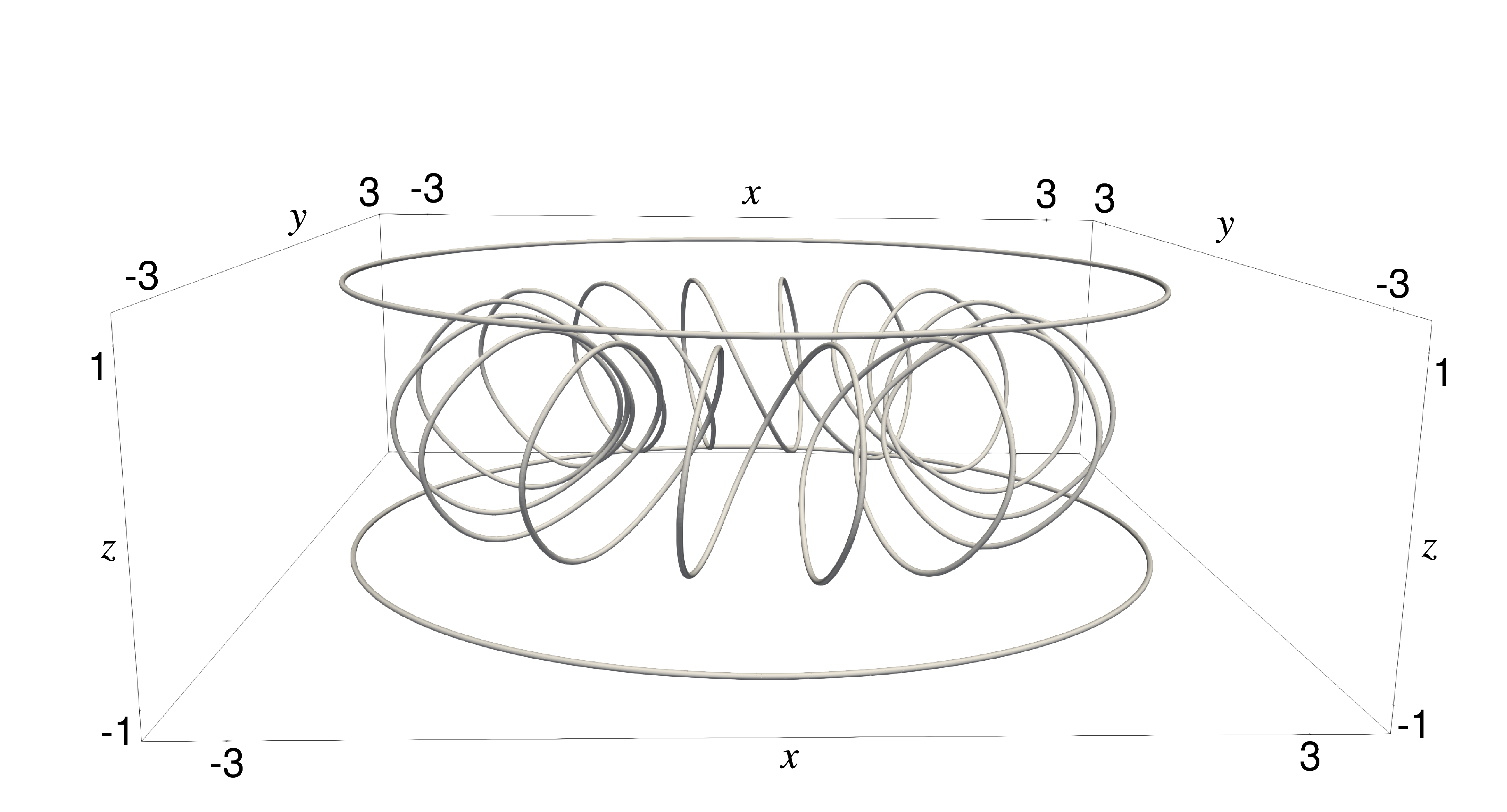}
    &\includegraphics[width=.5\textwidth, valign=t]{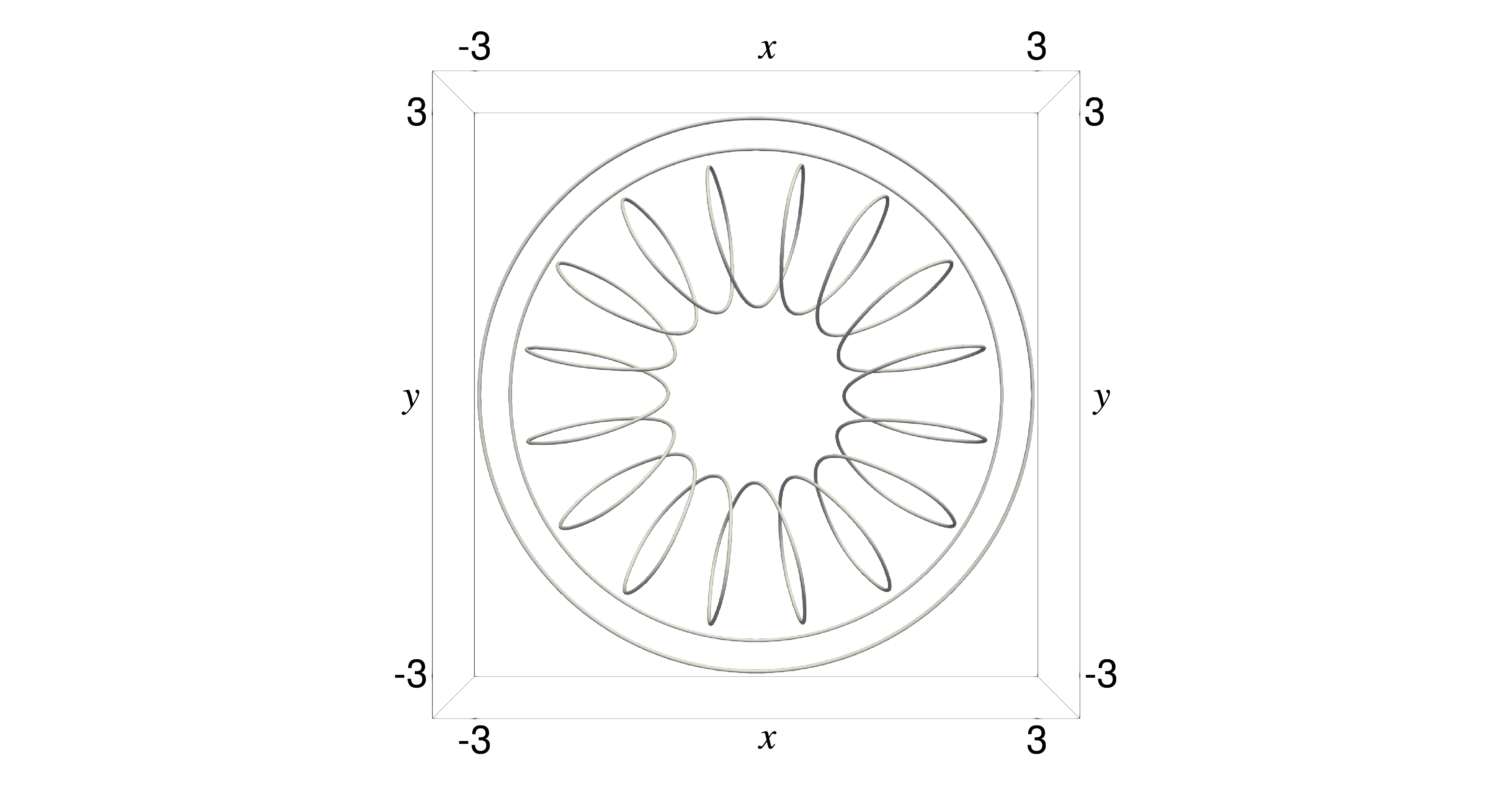}\\ 
    Optimised shape\newline\mbox{(Case III)}
    &\includegraphics[width=.4\textwidth, valign=t]{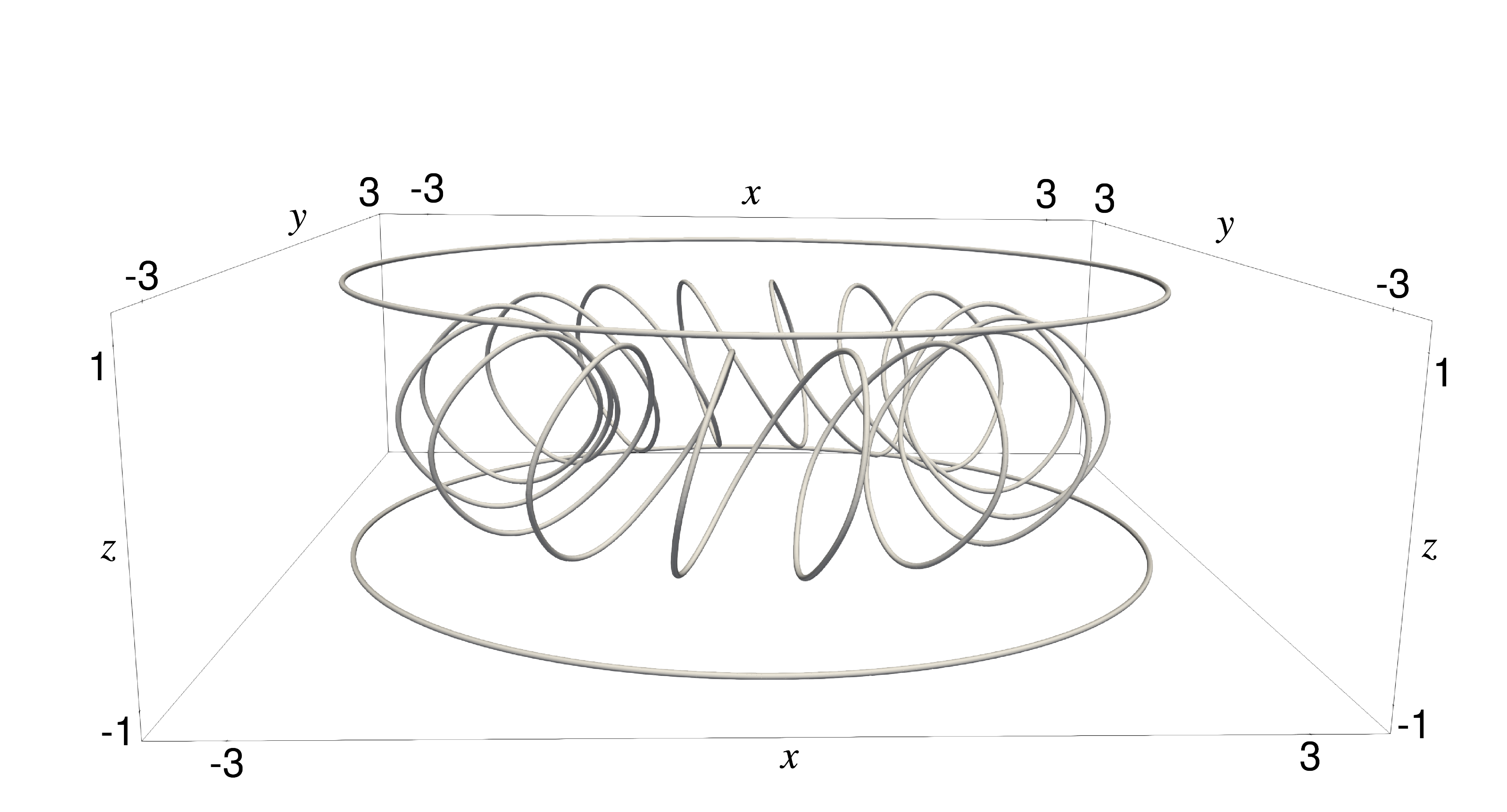}
    &\includegraphics[width=.5\textwidth, valign=t]{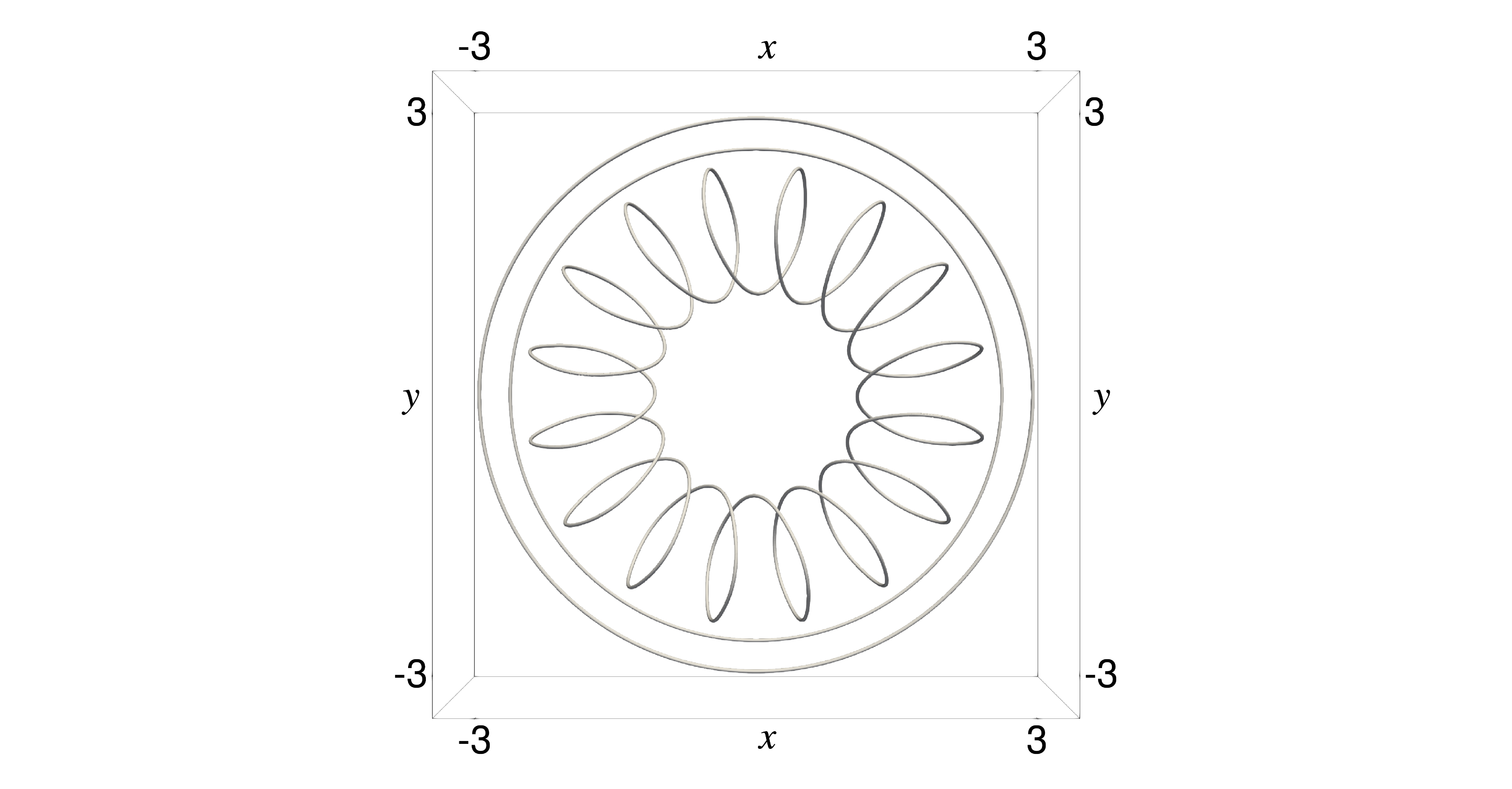}
  \end{tabular}
  \caption{Example 3: Initial and optimised shapes. For the reference, a bounding box of $[-3.45,3.45]\otimes[-3.45,3.45]\otimes[-1.15,1.15]$ is shown in each subfigure.}
  \label{fig:toroidal-shape}
\end{figure}

Figure~\ref{fig:toroidal-shape} shows the optimised shapes in comparison with the initial shape, which is common to all the cases. It is observed that each turn was deflected in the direction of the magnetic field (i.e., counter-clockwise when viewed from the positive $z$-axis). Further, the resulting shapes achieved certain symmetries, which are not assumed explicitly. Especially, Case III, where the length of the toroidal coil is constrained, resulted in a rotational symmetry with respect to the $z$-axis. Meanwhile, Cases I and II obtained a different type of symmetry due to the constraints regarding the coordinates of CPs.

\begin{figure}[hbt]
  \centering
  \begin{tabular}{cc}
    \multicolumn{2}{c}{Initial shape}\\
    \includegraphics[width=.35\textwidth]{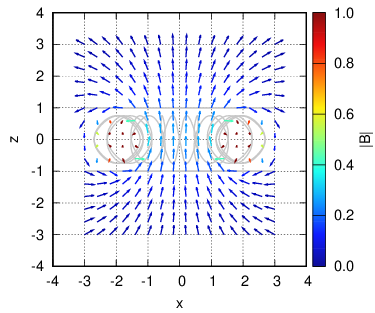}
    &\includegraphics[width=.35\textwidth]{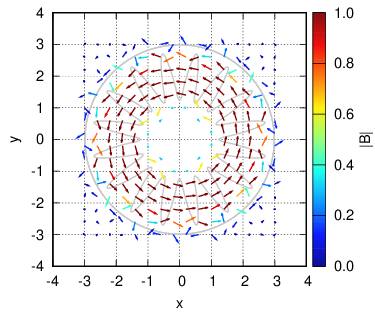}\\
    \multicolumn{2}{c}{Optimised shape (Case I)}\\
    \includegraphics[width=.35\textwidth]{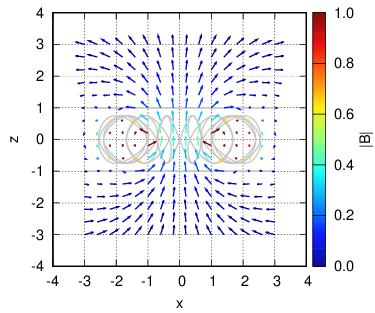}
    &\includegraphics[width=.35\textwidth]{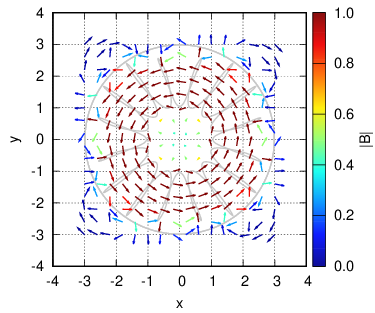}\\
    \multicolumn{2}{c}{Optimised shape (Case II)}\\
    \includegraphics[width=.35\textwidth]{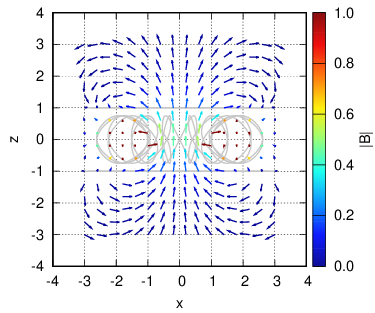}
    &\includegraphics[width=.35\textwidth]{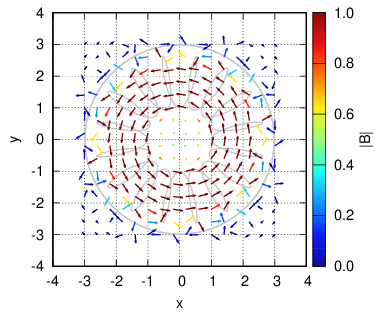}\\
    \multicolumn{2}{c}{Optimised shape (Case III)}\\
    \includegraphics[width=.35\textwidth]{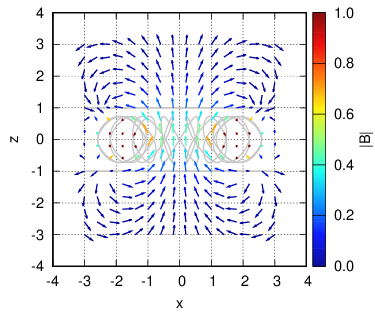}
    &\includegraphics[width=.35\textwidth]{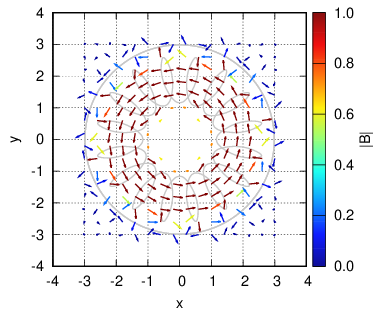}
  \end{tabular}
  \caption{Example 3: Distribution of the magnetostatic field $\bm{B}$ on the plane $y=0.0$ (left) and $z=0.0$ (right). The maximum of $|\bm{B}|$ is truncated to $1.0$.}
  \label{fig:toroidal-B}
\end{figure}

Finally, Figure~\ref{fig:toroidal-B} plots the vector field $\bm{B}$ on the planes $y=0.0$ and $z=0.0$. From the result of the former plane, it is observed that the magnetic field was more attracted to the toroidal coil $C^{(1)}$ in Cases II and III. However, the field on $z=0.0$ shows that the field inside $C^{(1)}$ was not particularly reinforced or confined by the optimisation even in those cases.

In order to assess the magnetic confinement more precisely, it is necessary to take account of the strength of $\bm{B}$ inside the toroidal coil into the objective function, according to the previous study~\cite{takahashi2024MRI}, for example. \footnote{Another idea is to maximise the self-inductance of $C^{(1)}$. However, such a self-inductance involves a diverging contour-integral and thus unavailable as an objective function in the wire model employed by this study.} The resulting SO is a multi-objective optimisation but out of scope of the present article.

Although the present example is not practical but rather mathematical, it successfully demonstrated the ability of the proposed SO system to handle a large and complicated problem.

\section{Conclusion}\label{s:conclusion}

This study presents a shape optimisation (SO) framework to achieve desired mutual inductances (MIs) between coils in three-dimensional space by manipulating their geometries. The key theoretical contribution lies in the derivation of the shape derivative of the objective function, defined as the deviation from the target MIs. To accommodate arbitrary coil shapes, the coils are presented by closed B-spline curves, following the approach of Takahashi et al. \cite{takahashi2024MRI}. Consequently, the shape of each coil can be designed by its control points (CPs). Furthermore,  a constraint on the coil length is formulated in terms of these CPs. The resulting SO system enables us to solve an SO problem using a quasi-Newton method, treating the discretised shape derivative as the gradient of the objective function.

In Section \ref{s:num}, numerical experiments were conducted to quantitatively and qualitatively validate the theoretical and numerical developments of the proposed SO system. However, the application of this system to more practical optimisation scenarios, including multi-objective optimisations involving MIs, remains a subject for future investigation.

\section*{Acknowledgements}

The authors express their gratitude to the anonymous referee for their constructive feedback, which greatly enhanced the quality of this manuscript.

\section*{Data Availability Statement}

The software and the associated data that support the findings of this study are available from the corresponding author upon reasonable request.

\section*{Declaration of generative AI and AI-assisted technologies in the writing process}

During the preparation of this work the authors used Google's Gemini (2.0 Flash) in order to improve language and readability. After using this tool/service, the authors reviewed and edited the content as needed and take full responsibility for the content of the publication.



\end{document}